\documentstyle[aps,prb,twocolumn,epsf,floats]{revtex}
\begin{document}

\ifx\hyperlink\hyperudefined
\else
\errmessage{Looks ugly with hyperlinks.}
\stop
\fi

\draft

\title{Cu-Au, Ag-Au, Cu-Ag and Ni-Au intermetallics: \\
First-principles study of phase diagrams and structures}
\author{V. Ozoli\c{n}\v{s}, C. Wolverton, and Alex Zunger}
\address{ National Renewable Energy Laboratory, Golden, CO 80401}
\date{September 14, 1997}
\maketitle

{\let\clearpage\relax
\twocolumn[
\widetext\leftskip=0.1075\textwidth \rightskip\leftskip
\begin{abstract}
The classic metallurgical systems -- noble metal alloys -- that have
formed the benchmark for various alloy theories, are revisited.
First-principles fully relaxed general potential LAPW total energies
of a few ordered structures are used as input to a mixed-space
cluster expansion calculation to study the phase
stability, thermodynamic properties and bond lengths 
in Cu-Au, Ag-Au, Cu-Ag and Ni-Au alloys.
(i) Our theoretical calculations
correctly reproduce the tendencies of Ag-Au and Cu-Au to form
compounds and Ni-Au and Cu-Ag to phase separate at $T=0$~K.
(ii) Of all possible structures, Cu$_3$Au $(L1_2)$ and CuAu $(L1_0)$
are found to be the most stable low-temperature phases of
Cu$_{1-x}$Au$_x$ with transition temperatures of $530$~K and $660$~K,
respectively, compared to the experimental values $663$~K and 
$\approx 670$~K. The significant improvement over previous
first-principles studies is attributed to the more 
accurate treatment of atomic relaxations in the present work.
(iii) LAPW formation enthalpies demonstrate that $L1_2$,
the commonly assumed stable phase of CuAu$_3$, is {\it not\/} 
the ground state for Au-rich alloys, but rather that ordered
$\langle 100 \rangle$ superlattices are stabilized.
(iv) We extract the
non-configurational (e.g., vibrational) entropies of formation and
obtain large values for the size mismatched systems:
0.48~$k_B$/atom in Ni$_{0.5}$Au$_{0.5}$ ($T=1100$~K),
0.37~$k_B$/atom in Cu$_{0.141}$Ag$_{0.859}$ ($T=1052$~K), 
and 0.16~$k_B$/atom in Cu$_{0.5}$Au$_{0.5}$ ($T=800$~K).
(v) Using 8~atom/cell special quasirandom structures we study
the bond lengths in disordered Cu-Au and Ni-Au alloys and
obtain good qualitative agreement with recent EXAFS measurements.
\end{abstract}
]}

\narrowtext

\section{Introduction: Chemical trends in noble metal alloys}
\label{sec:intro}

Noble metal alloys are, experimentally, among the most studied
intermetallic
systems.\cite{cuau_phase_D,hultgren,hansen,massalski,book_noble,cowley50/1,ogawa51,batterman57,kuczynski,kubiak,dheurle61,orr60/2,oriani54,oriani57,orr60/1,hirabayashi57,bienzle92,bienzle95,claeson84,sohal89,nahm95,shinohara96,exafs-niau,frenkel}
In addition, the Cu-Au system has been considered the classic paradigm
system for applying
different theoretical techniques of phase diagram and phase stability
calculations.\cite{shockley38,cowley50/2,baal73,golosov73,kikuchi74,didier77,kikuchi+sanchez,sigli+sanchez,kikuchi86,oates-cuau,wu71,chakrab92,chiolero,chakraborty95,ackland,Nbodypot,BSF,mazzone,seok97,guima95,guima-97,ahlers,clerirosato,terakura,sanchez+moruzzi,amador,wei87,lu-big,zwlu-niau,zwlu-agau,zwlu-SLs,paxton,LASTO88,weinb94,duane93,kudrnovsky91,abrikosov91,ruban95,ginatempo88}
 Most notably, this system has been
considered as the basic test case for the classic Ising-hamiltonian
statistical-mechanics treatment of
alloys.\cite{shockley38,cowley50/2,baal73,golosov73,kikuchi74,didier77,kikuchi+sanchez,sigli+sanchez}
More recently, noble metal binary alloys have been treated
theoretically via empirical fitting of the constants in Ising
hamiltonians,\cite{shockley38,cowley50/2,baal73,golosov73,kikuchi74,didier77,kikuchi+sanchez,sigli+sanchez,kikuchi86,oates-cuau}
semiempirical interatomic
potentials,\cite{wu71,chakrab92,chiolero,chakraborty95,ackland,Nbodypot,BSF,mazzone,seok97,guima95,guima-97,ahlers,clerirosato}
and via first-principles cluster
expansions.\cite{terakura,sanchez+moruzzi,amador,wei87,lu-big,zwlu-niau,zwlu-agau,zwlu-SLs}
The essential difference in philosophy between the classical
application of Ising 
models to
CuAu\cite{shockley38,cowley50/2,baal73,golosov73,kikuchi74,didier77,kikuchi86}
and more modern approaches 
based on the density functional formalism\cite{DF} is that in the former
approach the range and magnitudes of the interactions are postulated at
the outset (e.g., first or second neighbor pair interactions),
while the latter approaches make an effort to determine the
interactions from an electronic structure theory. However, despite recent
attempts,\cite{terakura,sanchez+moruzzi,amador,wei87,lu-big,zwlu-niau,zwlu-agau}
it is still not clear whether the noble metal alloys can be
essentially characterized as systems with short-range pair
interactions, or not.

\begin{table*}[th]
\caption{Major physical properties of Ag-Au, Cu-Ag, Cu-Au and Ni-Au
alloys. We give constituent size
mismatches, $\Delta a/\overline{a} = 2 (a_{\rm A} - a_{\rm B}) / 
(a_{\rm A} + a_{\rm B})$, electronegativity differences on the
Pauling scale,\protect\cite{pauling} $\Delta \chi$, mixing enthalpies of
the disordered alloys at the equiatomic composition, $\Delta H_{\rm
mix}(x=\frac{1}{2})$, signs of the nearest-neighbor Ising interaction,
$J_2$, order-disorder transition temperatures (or miscibility gap
temperatures for Cu-Ag and Ni-Au), $T_c(x=\frac{1}{2})$, and excess
entropies of solid solutions, $\Delta S^{\rm form}_{\rm tot} - \Delta
S_{\rm ideal}$. All phases are fcc-based.}
\begin{tabular}{ccccccccc}
System & $\Delta a/\overline{a}$\tablenotemark[1] & $\Delta
 \chi$\tablenotemark[2] & $\Delta H_{\rm mix}(x=1/2) $ & $J_2$ & Low-T
 phases\tablenotemark[7] & $T_c(x=\frac{1}{2}) $ 
 & $\Delta S^{\rm form}_{\rm tot} - \Delta S_{\rm ideal}$\tablenotemark[7]\\
& & & (meV/atom) & & & (K) & ($k_B$/atom)\\
\tableline
Cu-Au & $12\%$ & 0.64 & $-91$\tablenotemark[3] & $>0$ & 
$L1_2$, $L1_0$, $L1_2$(?) & 683\tablenotemark[7] & $+0.36$ \\
Ag-Au & $ 0\%$ & 0.61 & $-48$\tablenotemark[4] & $>0$ & $L1_2$,
 $L1_0$, $L1_2$& 115-168\tablenotemark[8] & $-0.17$ \\
Cu-Ag & $12\%$ & 0.03 & $+80$\tablenotemark[5] & $<0$ & 
Phase sep.     & $>T_m$ & $+0.04$ \\
Ni-Au & $15\%$ & 0.63 & $+76$\tablenotemark[6] & $>0$ & Phase sep.&
 1083\tablenotemark[4] & $+0.35$ \\
\end{tabular}
\tablenotetext[1]{Ref.~\protect\onlinecite{pearson}.}
\tablenotetext[2]{Ref.~\protect\onlinecite{pauling}.}
\tablenotetext[3]{Refs.~\protect\onlinecite{orr60/1,oriani57,hultgren}.}
\tablenotetext[4]{Ref.~\protect\onlinecite{hultgren}.}
\tablenotetext[5]{Theoretically calculated value from this work.}
\tablenotetext[6]{Refs.~\protect\onlinecite{hultgren,bienzle95}.}
\tablenotetext[7]{Refs.~\protect\onlinecite{hultgren,massalski}.}
\tablenotetext[8]{Ref.~\protect\onlinecite{AgAu-lowT}.}
\label{tab:physprop}
\end{table*}

Now that first-principles cluster expansion
approaches\cite{dblaks92,NATO} have 
advanced to the stage where both $T=0$ ground state structures and
finite-temperature thermodynamic quantities can be predicted without
any empirical information, it is interesting to take a {\it global
look\/} at the noble metal alloy family. Table~\ref{tab:physprop}
summarizes some of the salient
features\cite{cuau_phase_D,hultgren,hansen,massalski,oriani57,orr60/1,bienzle95,pearson,pauling,AgAu-lowT}
of the four binary systems Cu-Au, Ag-Au, Cu-Ag and Ni-Au. We included
the relative lattice constant mismatch $\Delta a / \overline{a}
= 2 \left| a_A - a_B \right| / \left| a_A + a_B\right|$ between the
consituents,\cite{pearson} the electronegativity difference $\Delta \chi
= \chi_A - \chi_B$ on the Pauling scale,\cite{pauling} the mixing
enthalpy of the equiatomic
alloy,\cite{hultgren,bienzle95} the sign of the calculated 
nearest neighbor pair interaction $J_2$ (present study), the
structural identity of the low-temperature
phases\cite{cuau_phase_D,hultgren,hansen,massalski,pearson} and the
order-disorder transition (or miscibility gap) 
temperatures\cite{hultgren,AgAu-lowT} $T_c$. Some interesting
observations and trends which we will attempt to reproduce
theoretically, are apparent from this general survey:

(i) Despite a large (12\%) size mismatch in Cu-Au,
and a small ($\approx 0$\%) size mismatch in Ag-Au, {\it both\/}
systems form ordered compounds at low temperatures and have negative
mixing enthalpies, suggesting attractive (``antiferromagnetic'')
A--B interactions. Thus, when the difference in the electronegativity
$\Delta \chi$ of the constituent atoms is sufficiently large, as it is
in CuAu and AgAu, size mismatch apparently does not determine ordering
{\it vs.\/} phase separation tendencies.

(ii) Despite a similar size mismatch (12\%) in Cu-Au and Cu-Ag, the
former orders while the latter phase-separates. Thus, the existence of
large electronegativity difference in Cu-Au (as opposed to the small
difference in Cu-Ag), seems to induce ordering tendencies.

(iii) Cu-Ag and Ni-Au both phase-separate (and have positive $\Delta
H_{\rm mix}$) as they have large size mismatches. Yet, Ni-Au having a
large electronegativity difference, shows an ordering-type
nearest-neighbor pair interaction ($J_2>0$), just like the compound
forming Cu-Au and Ag-Au, while Cu-Ag has a clustering-type
nearest-neighbor interaction ($J_2<0$). Thus, the sign of $J_2$ does
not reflect the low temperature ordering {\it vs.}
phase separation.

(iv) The amount $\Delta S_{\rm XS} = \Delta S_{\rm tot}^{\rm expt} -
\Delta S_{\rm ideal}$ by which the measured entropy\cite{hultgren}
$\Delta S^{\rm expt}_{\rm tot}$ deviates from the 
ideal configurational entropy $\Delta S_{\rm ideal} = 
k_B \left[ x \log x + (1-x) \log (1-x) \right]$, is unexpectedly 
large in Cu-Ag and Ni-Au, indicating a large non-configurational
entropy of formation.

Other interesting facts about the noble metal binary intermetallics
include:

(v) Despite numerous
studies,\cite{cuau_phase_D,hultgren,hansen,massalski,ogawa51,batterman57,kubiak,dheurle61,orr60/2}
the structure of the ordered phases in Au-rich Cu-Au is not well
established yet. It is 
often assumed\cite{cuau_phase_D,hultgren,hansen,massalski} that the
stable Au-rich  
low-temperature phase is CuAu$_3$ in the $L1_2$ structure, but direct
experiments\cite{ogawa51,batterman57,kubiak} below the 
order-disorder transition temperature $T_c (x=\frac{3}{4}) \approx
500$~K are difficult because the diffusion rates are very low and 
even the best ordered samples contain significant disorder.
Possible further thermodynamic transformations at lower temperatures
may be kinetically inhibited.

(vi) The trends in bond lengths {\it vs.} composition 
are non-trivial. Traditionally, all coherent-potential-approximation
based theories\cite{CPA-orig,CPA-reviews,ducastelle} of intermetallic
alloys have assumed that the nearest-neighbor bond lengths are equal,
$R_{AA}=R_{AB}=R_{BB}$, and proportional to the average lattice
constant. Recent theories\cite{zwlu-cu3pd,thorpe,papanik}
suggested, however, that bond lengths relax in the alloy to new
values, and this has a significant effect on the electronic
structure.\cite{zwlu-niau,zwlu-agpd+agau,zwlu-nipt} Recent EXAFS
experiments on NiAu\cite{exafs-niau} and CuAu\cite{frenkel} show
distinct  $R_{AA} \neq R_{AB} \neq R_{BB}$ bond lengths, which need to
be explained.

In this work we will analyze the above mentioned trends in terms
of a first-principles mixed-space cluster
expansion,\cite{dblaks92,NATO} based on modern local density
approximation (LDA) total energy calculations. We reproduce the
observed trends (i)-(vi) in ordering
preferences, mixing enthalpies $\Delta H_{\rm mix}$, transition
temperatures $T_c$ and interatomic bond lengths. In addition, we
predict new, hitherto unsuspected ordered phases in Au-rich Cu-Au
alloys.

\section{Basic ideology and methodology}
\label{sec:CE}

There are many problems in solid state physics that require knowledge
of the total energy $E(\sigma)$ of a lattice with $N$ sites as a
function of the occupation pattern $\sigma$ of these sites by atoms 
of types $A$ and $B$. This information is needed, for example,
in the ground state search problem,\cite{ducastelle} where one seeks
the configuration with the lowest energy at $T=0$~K. $\{ E(\sigma) \}$
is also needed for calculating the temperature- and composition-dependent
thermodynamic functions and phase diagrams of an $A_{1-x}B_x$ alloy.

A direct, quantum-mechanical calculation
of the total energy $E_{\rm direct} (\sigma) = \langle \Psi | \hat{H}
| \Psi \rangle / \langle \Psi | \Psi \rangle $ (where $\Psi$ is the
electronic ground state wave function and $\hat{H}$ is the
many-electron Hamiltonian) is possible only for a limited set of
configurations $\sigma$. This is so because (i) the computational
effort to solve the 
Schr\"odinger equation for a single configuration scales as the cube
of the number of atoms per unit cell, so that only small 
unit cells can be considered, (ii) there are $2^N$ configurations, and
(iii) for each configuration, one has to find the atomic relaxations
$\delta {\bf u}_{\rm min} (\sigma)$ which minimize the total energy.
Consequently, one searches for a ``cluster expansion'' (CE) that
accurately reproduces the results of a direct, quantum-mechanical
(e.g., LDA) calculation
\begin{equation}
\label{eq:eq1}
E_{\rm CE} (\sigma) \cong E_{\rm direct} (\sigma),
\end{equation}
without the unfavorable scaling of the computational cost with the
size of the unit cell. 

In designing a cluster expansion, there are few choices of independent
parameters. For example, one could choose to obtain a cluster expansion
for the volume- ($V$) dependent equation of state $E_{\rm direct}
(\sigma, V)$ [see, e.g.,
Refs.~\onlinecite{lu-big,connolly,didier-review}], or to
find a cluster expansion for the energy at the volume $V_{\rm
min}(\sigma)$ that minimizes $E_{\rm direct} (\sigma,V)$. We choose the
latter possibility. Furthermore, for each configuration $\sigma$, we
wish to reproduce the total energy corresponding to the fully relaxed
cell shape and atomic positions 
$\{ \delta {\bf u}_{\rm min} (\sigma) \}$. In other words, we choose to
represent
\begin{equation}
\label{eq:eq2}
E_{\rm CE} (\sigma) \cong E_{\rm direct} [\sigma; \delta {\bf u}_{\rm
min} (\sigma); V_{\rm min} (\sigma) ] \equiv E_{\rm direct} (\sigma).
\end{equation}
Note that by focusing on the equilibrium energy of each configuration,
we give up the possibility of studying non-equilibrium geometries
(e.g., bond lengths) and equations of state. Instead, for each
occupation pattern $\sigma$, we can find the total energy $E(\sigma)$
of the atomically relaxed and volume-optimized geometry.

The best-known cluster expansion is the generalized Ising model in
which the equilibrium total energy of an {\it arbitrary\/} 
configuration $\sigma$ is expanded in a series
of basis functions defined as pseudospin products on the crystal
sites:
\begin{eqnarray}
\nonumber
E(\sigma) = J_0 + \sum_i J_i S_i + \frac{1}{2} \sum_{i \ne j} J_{ij}
S_i S_j \\
\label{eq:basicCE}
+ \frac{1}{3!} \sum_{i \ne j \ne k} J_{ijk} S_i S_j S_k + \ldots,
\end{eqnarray}
where in binary $A_{1-x}B_x$ alloys $S_i = +1$ or $-1$,
depending on whether the site $i$ is occupied by an atom of type $A$
or $B$. Equation~(\ref{eq:basicCE}) is valid whether the lattice is
relaxed ot not, as long as a one-to-one correspondence exists between
the actual positions of atoms and the ideal fcc sites.
The practical usefulness of the cluster
expansion Eq.~(\ref{eq:basicCE}) rests on the assumption that the
effective cluster interactions (ECI's), $J_{ij}, J_{ijk}, \ldots,$
are rapidly decreasing functions of the number of sites and intersite
separation, so that only a finite number of terms has to be kept
in Eq.~(\ref{eq:basicCE}) for the desired accuracy. In this
case, we can write the formation enthalpy of structure $\sigma$,
\begin{equation}
\label{eq:LDAenergies}
\Delta H_{\rm direct} (\sigma) = E(\sigma) - x E_A - (1-x) E_B,
\end{equation}
where $E_A$ and $E_B$ are total energies of the pure constituents $A$
and $B$, as the following cluster expansion (CE):
\begin{equation}
\label{eq:truncCE}
\Delta H_{\rm CE} (\sigma) = J_0 + \sum_f^{N_f} D_f J_f
\overline{\Pi}_f (\sigma).
\end{equation}
Here $N_f$ is the number of nonzero effective interactions and
$\overline{\Pi}_f (\sigma)$ are lattice averages of the spin products
in configuration $\sigma$. 

Sanchez, Ducastelle and Gratias\cite{sanchez84} have shown
that there is a set of composition-independent
interactions for Eq.~(\ref{eq:basicCE}) which can
exactly reproduce the directly calculated total energies of {\it all\/}
configurations $\sigma$.
This statement is strictly true if all possible clusters are included
in Eq.~(\ref{eq:basicCE}), and should hold for the truncated series
Eq.~(\ref{eq:basicCE}) if the cluster expansion is well converged.
Several methods\cite{asta91,sanchez93} yield concentration-dependent
effective interactions, providing in principle equally valid
schemes for representing $\Delta H_{\rm direct}(\sigma)$ in
terms of a cluster expansion. In the present work, we select
composition-independent interactions, since these can be directly
compared to previous Ising model
treatments\cite{shockley38,cowley50/2,baal73,golosov73,kikuchi74,didier77,kikuchi+sanchez,sigli+sanchez,kikuchi86,oates-cuau,terakura,sanchez+moruzzi,amador,wei87,lu-big,zwlu-niau,zwlu-agau,zwlu-SLs}
of the noble metal alloy phase diagrams.

A number of issues arise in trying to construct a cluster expansion
that satisfies Eq.~(\ref{eq:eq2}):

(i) The number of interactions and their types (pair, multibody)
cannot be decided arbitrarily, but must be constrained by a
microscopic electronic-structure theory according to
Eqs.~(\ref{eq:eq1}) and (\ref{eq:eq2}).

(ii) In most configurations $\sigma$, atoms move away from the ideal
lattice sites, which not only lowers the total energies $E_{\rm direct}
(\sigma)$, but also slows down the convergence\cite{italians} of the
expansion Eq.~(\ref{eq:basicCE}). The solution is to have a cluster
expansion with many interaction terms $N_J$ that can represent such
situations. We accomplish this by using a reciprocal space
formulation, formally equivalent to an infinite number of real-space 
pair interactions. 

(iii) Some cluster expansions\cite{connolly} require that the number
of interactions $N_J$ must equal the number of configurations $N_\sigma$
whose total energies need to be evaluated via the direct
electronic-structure method. The number
of such calculations may be excessive in view of (ii). We thus
introduce a method in which $N_\sigma \ll N_J$. Furthermore,
interactions that are not needed to satisfy Eq.~(\ref{eq:eq2})
are automatically discarded.

(iv) One has to deal with the macroscopic elastic strain leading to a
${\bf k} \rightarrow 0$ singularity in the Fourier transform of the
pair interactions,
\begin{equation}
\label{eq:jofk}
J_{\rm pair} ({\bf k}) = \sum_j D_j J_{\rm pair} ({\bf R}_i-{\bf R}_j)
e^{-i{\bf kR}_j},
\end{equation}
where $D_j$ is the number of $\{{\bf R}_i, {\bf R}_j\}$ pairs per
lattice site. As shown by Laks {\it et al.\/}\cite{dblaks92} (see also
the discussion below), in size mismatched systems the correct
$J_{\rm pair} ({\bf k})$ depends on direction $\hat{k}$ in the
long-wavelength limit ${\bf k} \rightarrow 0$. To solve this,
we express $J_{\rm pair} ({\bf k})$ as a sum of two parts,
\begin{equation}
\label{eq:Jsplit}
J_{\rm pair} ({\bf k}) = J_{\rm SR} ({\bf k}) + J_{\rm CS} (\hat{k}),
\end{equation}
where $J_{\rm SR} ({\bf k})$ is an analytic function of {\bf k} and
can be obtained from short-ranged real space pair interactions, while
$J_{\rm CS} (\hat{k})$ contains the nonanalytic behavior around
${\bf k}=0$ and depends only on the direction $\hat{k}$.
To explain this singularity, we consider the energy of a coherent
$A_nB_n$ superlattice, formed by a periodic stacking of $n$ layers of
$A$ and $n$ layers of $B$ in direction $\widehat{G}$.
By introducing the structure factor,
\begin{equation}
\label{eq:Sofk}
S({\bf k}, \sigma) = \sum_j S_j e^{-i{\bf kR}_j},
\end{equation}
the total pair interaction energy in Eq.~(\ref{eq:basicCE}) 
can be expressed as a reciprocal space sum:
\begin{equation}
\label{eq:Epair}
E_{\rm pair} (\sigma) = \sum_{\bf k} J_{\rm pair} ({\bf k})
\, \left| S({\bf k},\sigma) \right|^2.
\end{equation}
$A_nB_n$ superlattice has nonzero
structure factor at ${\bf k} = \frac{1}{2n} \widehat{G}$, and
therefore its energy is determined by $J_{\rm pair} (\frac{1}{2n}
\widehat{G})$. As $n \rightarrow \infty$
its formation energy is given by a sum of the
epitaxial deformation energies of pure constituents needed to bring
them to a common lattice constant in the plane perpendicular to
$\widehat{G}$. Since the epitaxial deformation energy of
pure constituents is direction dependent (e.g., it is easier to
stretch Cu in [100] planes than in [111] planes, see
Sec.~\ref{sec:Ecs}), the formation energy $\Delta H(A_\infty B_\infty)$
is also direction dependent. Therefore, $\lim_{{\bf k}
\rightarrow 0} J_{\rm pair} ({\bf k})$ must depend on the direction of
approach to the origin, proving that $J_{\rm pair} ({\bf k})$ is
singular. Physically, the nonanaliticity of $J_{\rm pair} ({\bf k})$
is caused by long-range interactions via macroscopic elastic
strain and cannot be reproduced using finite-ranged real-space pair
interactions, but must be accounted for explicitly in reciprocal
space. If the singularity is neglected, then as explained in
Ref.~\onlinecite{dblaks92}, the cluster expansion fails
not only for long-period ($n \rightarrow \infty$) superlattices
$A_nB_n$, but also for those short-period ($n>2$) superlattices which
have not been explicitly included in the constraint
Eq.~(\ref{eq:eq2}).
We emphasize that although the contribution of $J_{\rm sing} ({\bf
k})$ to the formation energy is nonzero only in size-mismatched
systems, it is not related to the atomic relaxation energy
for a particular structure $\sigma$ in any simple way (except if
$\sigma$ itself is a long-period superlattice).

The singularity in $J_{\rm pair}({\bf k})$ is similar to the
singularity in the dynamical matrix 
$D_{\alpha\beta} (\kappa\kappa' | {\bf k})$ of polar crystals 
in the long-wavelength limit,\cite{born} caused by long-range
electrostatic interactions via macroscopic electric field. 
In lattice dynamics, $D_{\alpha\beta} (\kappa\kappa' | {\bf k})$
is expressed as a sum of regular and singular parts,
$D_{\alpha\beta} (\kappa\kappa' | {\bf k}) =
D^{\rm sing}_{\alpha\beta} (\kappa\kappa' | {\bf k}) + 
D^{\rm reg}_{\alpha\beta} (\kappa\kappa' | {\bf k})$, where
$D^{\rm reg}_{\alpha\beta} (\kappa\kappa' | {\bf k})$ (analytic
as ${\bf k} \rightarrow 0$) is due to short-range force constants.
The singular part  $D^{\rm sing}_{\alpha\beta} (\kappa\kappa' | {\bf
k})$ gives rise to LO/TO splitting of the zone-center optical
frequencies  $\omega_\Gamma$ in polar crystals, and also leads to
a directional dependence of $\omega_\Gamma (\hat{k})$ in uniaxial
crystals (e.g., CuPt-type GaInP$_2$). These phenomena cannot be
reproduced by any set of finite-ranged microscopic force constants, but
have to be calculated explicitly using the macroscopic Maxwell
equations.\cite{maradudin}

In summary, we seek to find a function $E_{\rm CE} (\sigma)$ which
accurately reproduces the LDA total energies $E_{\rm LDA}
[\sigma,\delta {\bf u}_{\rm min}(\sigma); V_{\rm min} (\sigma) ]
\equiv E_{\rm LDA} (\sigma)$ at the atomically relaxed geometry and
equilibrium volume of configuration $\sigma$. The function $E_{\rm CE}
(\sigma)$ we consider includes composition- and volume-independent
interactions, so as to maintain maximum similarity with the classical
Ising model. The number and type of interactions are not decided
arbitrarily, but are constrained by the electronic
structure theory  used (here, the LDA). Relaxation is treated 
accurately by including long-range pair interactions in the
reciprocal space representation. The ${\bf k} \rightarrow 0$
singularity, affecting both short and long-period superlattices, is
dealt with explicitly.

The above requirements are satisfied by the mixed space
cluster expansion (MSCE):
\begin{eqnarray}
\nonumber
\Delta H_{\rm CE} (\sigma) = \sum_{\bf k} J_{\rm pair} ({\bf k})
\, \left| S({\bf k},\sigma) \right|^2 \\
\label{eq:recipCE}
+ \sum_f^{\rm MB} D_f J_f \overline{\Pi}_f (\sigma)
+ \Delta E_{\rm CS} (\sigma).
\end{eqnarray}
We have separated out the so-called equilibrium constituent
strain energy term, $\Delta E_{\rm CS} (\sigma)$, which accounts for
the ${\bf k} \rightarrow 0$ singularity.\cite{dblaks92}
In Eq.~(\ref{eq:recipCE}) we do not need to
calculate $\Delta E_{\rm CS} (\sigma)$ for each configuration $\sigma$,
but only for the {\it directions\/} $\widehat{k}$ of the wave vectors
with nonzero $S({\bf k}, \sigma)$. In fact, it is constructed to
coincide with the elastic strain energy of coherent superlattices in
the long-period limit:\cite{dblaks92}
\begin{eqnarray}
\label{eq:Ecs(sigma)def}
\Delta E_{\rm CS} (\sigma) = \sum_{\bf k} J_{\rm CS} (x,\widehat{k})
\left| S({\bf k},\sigma) \right|^2,\\
\label{eq:Jcsdef}
J_{\rm CS} (x,\widehat{k}) = \frac{\Delta E^{\rm eq}_{\rm CS}(x,\widehat{k})}
{4x(1-x)},
\end{eqnarray}
where $S({\bf k},\sigma)$ is the structure factor from
Eq.~(\ref{eq:Sofk}). The quantity 
$\Delta E^{\rm eq}_{\rm CS}(x,\widehat{k})$ depends only on the
direction $\widehat{k}$, and will be given in Sec.~\ref{sec:Ecs}.
Equation~(\ref{eq:Ecs(sigma)def}) is 
{\it exact\/} for long-period superlattices, but 
represents a {\it choice\/} for short-period
superlattices and non-superlattice (e.g., $L1_2$) structures.
It has been found\cite{dblaks92} that the choice
Eq.~(\ref{eq:Ecs(sigma)def}) improves the cluster expansion
predictions also for short-period superlattices.

\begin{table*}[t!]
\caption{Definition of the small-unit-cell ordered structures used in
the LDA total energy calculations.}
\begin{tabular}{c|ccccc}
\multicolumn{6}{c}{\normalsize Simple superlattices}\\
\tableline
Compo- & \multicolumn{5}{c}{Orientation} \\
sition & (001) & (011) & (111) & (311) & (201) \\
\tableline
AB & $L1_0$ (CuAu) & $L1_0$ (CuAu) & $L1_1$ (CuPt) & $L1_1$ (CuPt) &  $L1_0$ (CuAu)\\
A$_2$B & ``$\beta1$'' (MoSi$_2$) & ``$\gamma1$'' (MoPt$_2$) &
``$\alpha1$'' (CdI$_2$)& ``$\gamma1$'' (MoPt$_2$) & ``$\gamma1$''(MoPt$_2$)\\
AB$_2$ & ``$\beta2$'' (MoSi$_2$) & ``$\gamma2$'' (MoPt$_2$) &
``$\alpha2$'' (CdI$_2$)& ``$\gamma2$'' (MoPt$_2$) & ``$\gamma2$''(MoPt$_2$)\\
A$_3$B & ``Z1''                  & ``Y1''                   & ``V1'' &  ``W1'' &$D0_{22}$ (TiAl$_3$)\\
AB$_3$ & ``Z3''                  & ``Y3''                   & ``V3'' &  ``W3'' &$D0_{22}$ (TiAl$_3$)\\
A$_2$B$_2$ & ``Z2''              & ``Y2''                   & ``V2'' &
``W2'' & ``40''(CuFeS$_2$)      \\
\tableline
\tableline
\multicolumn{6}{c}{\normalsize Other structures}\\
\tableline
Compo- & Name & Prototype & Superlattice & Period & Reference \\
sition &      &           & direction    &        &      \\
\tableline
A$_3$B$_1$ & $L1_2$   & Cu$_3$Au & none  &       &
\protect\onlinecite{lu-big}\\
A$_1$B$_3$ & $L1_2$   & Cu$_3$Au & none  &       &
\protect\onlinecite{lu-big}\\
A$_7$B   & $D7_a$     &          & none  &       &
\protect\onlinecite{lu-big}\\
A$_4$B$_4$ & $D4$     &          & none  &       &
\protect\onlinecite{lu-big}\\
AB$_7$ & $D7_b$       &          & none  &       &
\protect\onlinecite{lu-big}\\
A$_8$B &              & Ni$_8$Nb & none  &       &
\protect\onlinecite{villars}\\
AB$_8$ &              & Ni$_8$Nb & none  &       &
\protect\onlinecite{villars}\\
A$_6$B$_2$ & $D0_{23}$& Al$_3$Zr & (401) & (5,1,1,1) &
\protect\onlinecite{villars}\\
A$_6$B$_2$ & LPS-3    &          & (601) & (5,1,1,1) &
\protect\onlinecite{MCpaper}\\
A$_4$B$_4$ & SQS$8_{\rm a}$  &   & (311) & (2,3,2,1) &
\onlinecite{zwlu-agpd+agau}\\
A$_4$B$_4$ & SQS$8_{\rm b}$  &   & (311) & (3,2,1,2) &
\protect\onlinecite{zwlu-agpd+agau}\\
A$_6$B$_2$ & SQS$14_{\rm a}$ &   & (201) & (6,2)     &
\protect\onlinecite{zwlu-cu3pd}\\
A$_2$B$_6$ & SQS$14_{\rm b}$ &   & (201) & (2,6)     & 
\protect\onlinecite{zwlu-cu3pd}
\end{tabular}
\label{tab:strucs}
\end{table*}

Equation~(\ref{eq:recipCE}) is a generalized Ising model description of
the formation energy of {\it any\/} relaxed configuration $\sigma$,
even if a direct LDA calculation for this $\sigma$ is impractical.
The cluster interaction energies $\{ J_{\rm pair}({\bf k}) \}$ and 
$\{ J_f \}$ are obtained by fitting Eq.~(\ref{eq:recipCE}) to the LDA
formation energies. An additional 
smoothness requirement is imposed on $J_{\rm pair} ({\bf k})$, which
ensures that the pair interactions are optimally short-ranged in real
space. Namely, we minimize the sum
\begin{eqnarray}
\nonumber
\Delta_{\rm rms}^2 = \frac{1}{N_\sigma} \sum_\sigma w_\sigma
\left[ \Delta H_{\rm CE} (\sigma) - \Delta H_{\rm LDA} (\sigma)
\right]^2\\
\label{eq:CErms}
 + \frac{t}{\alpha} \sum_{\bf k} J_{\rm pair} ({\bf k})
\left[-\nabla^2_{\bf k} \right]^{\lambda/2} J_{\rm pair} ({\bf k}),
\end{eqnarray}
where $\lambda$ and $t$ are free parameters and $\alpha$ is
a normalization constant.\cite{dblaks92} Typically we choose
$\lambda=4$ and $t=1$, but the fit is not sensitive to this choice.

This approach solves the four problems indicated above in the sense
that (i) the fitting process itself
automatically selects the pair interactions that are essential to
obtain a good fit (process still does not select multibody figures),
(ii) the pair interactions can be of arbitrary long range,
facilitating treatment of systems with large elastic relaxations,
(iii) the number of pairs can be much larger than the number of
ordered structures in the fit, and
(iv) the directly calculated constituent strain energy
$\Delta E_{\rm CS}$ contains the ${\bf k} \rightarrow 0$ singularity.
Unlike all CPA-based methods,\cite{CPA-orig,CPA-reviews} the present
approach includes full account of atomic relaxation and local
environment effects. Unlike the classical Ising
descriptions,\cite{shockley38,baal73,golosov73,kikuchi74,didier77,kikuchi+sanchez,sigli+sanchez,kikuchi86}
the interaction energies are determined by the electronic structure
rather than being guessed. Finally, unlike the computational alchemy
linear response approach,\cite{italians} multibody terms are included
here.

Having written the expression for the total energy of arbitrary
configuration, Eq.~(\ref{eq:recipCE}), we can evaluate its constants
from a limited number of LDA calculations on small unit
cell ($N_{\rm atoms}<10$) ordered structures with fully relaxed
atomic positions. Equation~(\ref{eq:recipCE})
can then be employed in simulated annealing and Monte Carlo
calculations\cite{binder,MCpaper} yielding $T=0$ ground states and $T>0$
statistical and thermodynamic properties. Further details of the
method are given in Sec.~\ref{sec:details}.

\section{Details of the Method}
\label{sec:details}

\subsection{$T=0$ energetics}
\label{sec:LDA}

The calculations of $T=0$ total energies employ the full-potential
linearized augmented plane wave method\cite{lapw} (FLAPW). The basis
set consists of plane waves in the interstitial region, augmented in a
continuous and differentiable way with the solutions of the radial
Schr\"odinger equation inside the non-overlapping muffin-tin spheres.
Non-spherical potential and electronic charge density terms are
calculated in all space and included in the Hamiltonian matrix.
Core states are treated fully relativistically and recalculated in
each self-consistency iteration. The wave equation for the valence
states includes all relativistic effects except the spin-orbit
interaction, i.e., they are treated scalar relativistically.
FLAPW is the most accurate all-electron method,
superior to the methods employing overlapping atomic
spheres (atomic-spheres approximation -- ASA) and/or shape
approximations to the potential.

We use the Wigner exchange-correlation functional.\cite{wigner}
As a check, we have performed several
calculations using the Perdew-Zunger\cite{perdewzunger}
parametrization of the Ceperley-Alder\cite{ceperleyalder}
functional and the generalized gradient approximation of Perdew and
Wang.\cite{ggapw91} We find (see Sec.~\ref{sec:GS-CuAu}) that the 
various exchange-correlation functionals change the enthalpies
of formation of ordered Cu-Au compounds by a negligible
amount (less than 2~meV/atom).

\begin{table*}
\squeezetable
\caption{LDA calculated formation [Eq.~(\protect\ref{eq:LDAenergies})]
enthalpies for 
fcc superstructures (defined in Table~\protect\ref{tab:strucs}) 
of Ag-Au, Cu-Ag, Cu-Au and Ni-Au. The numbers in
parentheses represent errors of the cluster expansion fit. All
energies in meV/atom.} 
\begin{tabular}{cc|r|rr|rr|rr}
\multicolumn{2}{c|}{\normalsize Structure}&
\multicolumn{1}{c|}{\normalsize Ag-Au}&
\multicolumn{2}{c|}{\normalsize Cu-Ag}&
\multicolumn{2}{c}{\normalsize Cu-Au}&
\multicolumn{2}{c}{\normalsize Ni-Au}\\
&&&&&&&&\\
Superlattice & Name &
$\Delta H^{\rm LDA}_{\rm unrel}$ & 
$\Delta H^{\rm LDA}_{\rm unrel}$ & 
$\Delta H^{\rm LDA}_{\rm rel}$ &
$\Delta H^{\rm LDA}_{\rm unrel}$ & 
$\Delta H^{\rm LDA}_{\rm rel}$ &
$\Delta H^{\rm LDA}_{\rm unrel}$ & 
$\Delta H^{\rm LDA}_{\rm rel}$ \\
&&&&&&&&\\
\tableline

A & fcc &
$0.0\;(-0.4)$ & 0.0 & $0.0\;(-0.1)$ & 0.0 & $0.0\;(+0.2)$ & 0.0 & $0.0\;(+0.4)$\\
B & fcc & 
$0.0\;(-0.5)$ & 0.0 & $0.0\;(+0.3)$ & 0.0 & $0.0\;(-0.4)$ & 0.0 & $0.0\;(-0.2)$\\
&&&&&&&&\\

\underline{(001) Struct:}&&&&&&\\
A$_1$B$_1$ & $L1_0$ & 
$-59.7\;(-0.8)$ & $+107.6$ & $+100.5\;(+0.4)$ & $-36.1$ & $-48.2\;(+0.1)$ &
$+98.1$ & $+76.1\;(+1.4)$ \\
A$_2$B$_1$ & ``$\beta1$'' & 
$-40.8\;(-0.1)$ & $+130.2$ & $+90.8\;(-0.7)$ & $+51.0$ & $-3.8\;(-2.6)$ &
$+207.8$ & $+105.7\;(-0.1)$ \\
A$_1$B$_2$ & ``$\beta2$'' & 
$-40.0\;(+0.1)$ & $+112.0$ & $+75.0\;(+1.0)$ & $+40.1$ & $-40.8\;(+0.6)$ &
$+151.7$ & $+38.3\;(+0.1)$ \\
A$_3$B$_1$ & ``Z1'' & 
$-29.2\;(-0.1)$ & $+126.4$ & $+79.8\;(+1.8)$ & $+76.5$ & $+10.6\;(+0.3)$ &
$+221.7$ & $+89.9\;(-4.2)$ \\
A$_1$B$_3$ & ``Z3'' & 
$-27.9\;(+0.7)$ & $+96.8$ & $+56.9\;(-0.2)$ & $+50.0$ & $-28.2\;(+1.8)$ &
$+142.0$ & $+32.4\;(+4.0)$ \\
A$_2$B$_2$ & ``Z2'' & 
$-28.8\;(-0.3)$ & $+164.7$ & $+77.8\;(+0.4)$ & $+136.4$ & $-6.7\;(-1.0)$ &
$+286.7$ & $+70.2\;(+0.1)$ \\
A$_2$B$_3$ & ``Z5'' & 
& & & & & $+273.3$ & $+57.1\;(-0.8)$ \\
A$_3$B$_3$ & ``Z6'' &
& & & & & $+355.5$ & $+63.2\;(+0.7)$ \\
A$_\infty$B$_\infty$ &  & 
 $0.0\;(0.0)$ & & $+20.4\;(0.0)$ & & $+20.3\;(-0.1)$ & $+576.2$ & $+30.8\;(0.0)$ \\
&&&&&&&&\\

\underline{(111) Struct:}&&&&&&\\
A$_1$B$_1$ & $L1_1$ & 
$-43.0\;(-0.4)$ & $+134.8$ & $+129.8\;(-1.1)$ & $+60.3$ & $+32.5\;(-0.1)$ &
$+192.3$ & $+166.8\;(+1.4)$ \\
A$_2$B$_1$ & ``$\alpha1$'' & 
$-30.2\;(0.0)$ & $+152.4$ & $+120.4\;(-2.9)$ & $+123.0$ & $+61.4\;(-7.7)$ &
$+288.5$ & $+202.2\;(-6.4)$ \\
A$_1$B$_2$ & ``$\alpha2$'' & 
$-30.8\;(0.0)$ & $+124.9$ & $+95.0\;(+2.9)$ & $+86.4$ & $+2.1\;(+7.7)$ &
$+200.9$ & $+100.9\;(+6.4)$ \\
A$_3$B$_1$ & ``V1'' & 
$-21.3\;(+0.3)$ & $+145.9$ & $+108.4\;(+0.4)$ & $+136.1$ & $+78.6\;(+4.1)$ &
$+290.8$ & $+193.7\;(+4.1)$ \\
A$_1$B$_3$ & ``V3'' & 
$-21.4\;(+0.6)$ & $+106.8$ & $+73.6\;(+1.5)$ & $+79.5$ & $+5.1\;(+0.8)$ &
$+172.8$ & $+83.0\;(+4.0)$ \\
A$_2$B$_2$ & ``V2'' & 
$-22.9\;(-0.4)$ & $+177.1$ & $+109.1\;(-1.0)$ & $+170.6$ & $+52.2\;(-2.5)$ &
$+335.8$ & $+162.4\;(-4.1)$ \\
A$_\infty$B$_\infty$ &  & 
 $0.0\;(0.0)$ & & $+86.3\;(-1.0)$ & & $+95.8\;(+0.3)$ & $+576.2$ & $+173.8\;(+1.3)$\\
&&&&&&&&\\

\underline{(011) Struct:}&&&&&&\\
A$_2$B$_1$ & $\gamma1$ & 
$-49.7\;(-0.4)$ & $+106.4$ & $+100.3\;(-0.6)$ & $-14.2$ & $-18.4\;(+3.3)$ &
$+123.3$ & $+98.9\;(-3.8)$ \\
A$_1$B$_2$ & $\gamma2$ & 
$-46.9\;(+0.4)$ & $+97.2$ & $+92.5\;(+0.8)$ & $+1.7$ & $-6.7\;(-5.2)$ &
$+126.3$ & $+102.6\;(+3.8)$ \\
A$_3$B$_1$ & ``Y1'' &
$-37.0\;(0.0)$ & $+105.1$ & $+85.4\;(+3.5)$ & $+21.8$ & $-1.3\;(+3.8)$ &
$+148.5$ & $+99.2\;(+7.8)$ \\
A$_1$B$_3$ & ``Y3'' & 
$-35.4\;(+0.6)$ & $+85.5$ & $+75.2\;(-1.3)$ & $+19.4$ & $-1.0\;(+0.1)$ &
$+104.1$ & $+78.7\;(+1.1)$ \\
A$_2$B$_2$ & ``Y2'' & 
$-44.1\;(-0.3)$ & $+136.0$ & $+105.7\;(-1.1)$ & $+59.5$ & $-4.2\;(-2.0)$ &
$+192.3$ & $+96.6\;(-4.5)$ \\
A$_\infty$B$_\infty$ &  & 
 $0.0\;(0.0)$ & & $+75.3\;(-1.2)$ & & $+66.1\;(+0.3)$ & $+576.2$ & $+117.7\;(+1.6)$\\
&&&&&&&&\\

\underline{(113) Struct:}&&&&&&\\
A$_3$B$_1$ & ``W1''&
$-35.9\;(+0.5)$ & $+104.7$ & $+94.2\;(-0.2)$ & $+22.0$ & $+7.0\;(+1.5)$ &
$+125.7$ & $+120.8\;(+5.2)$ \\
A$_1$B$_3$ & ``W3''& 
$-34.4\;(-0.2)$ & $+98.6$ & $+91.4\;(+9.0)$ & $+21.1$ & $+7.8\;(+0.6)$ & &
$+88.4\;(+5.3)$ \\
A$_2$B$_2$ & ``W2''& 
$-50.6\;(-0.1)$ & $+121.9$ & $+104.7\;(-4.4)$ & $+15.7$ & $-20.9\;(-1.0)$ &
$+144.2$ & $+93.6\;(-5.3)$ \\
A$_\infty$B$_\infty$ &  & 
 $0.0\;(0.0)$ & & $+65.9\;(-1.4)$ & & $+69.5\;(+0.4)$ & $+576.2$ & $+119.8\;(+1.9)$\\
&&&&&&&&\\

\underline{(201) Struct:}&&&&&&\\
A$_3$B$_1$ & $D0_{22}$& 
$-42.3\;(-0.2)$ & $+85.2$ & $+85.1\;(+1.3)$ & $-32.7$ & $-32.8\;(+0.3)$ &
$+75.0$ & $+75.0\;(+5.6)$ \\
A$_1$B$_3$ & $D0_{22}$& 
$-41.0\;(-0.3)$ & $+76.8$ & $+76.4\;(-0.5)$ & $-10.6$ & $-11.8\;(-1.8)$ &
$+68.7$ & $+68.6\;(+1.5)$ \\
A$_2$B$_2$ & CH, or ``40''& 
$-55.3\;(+0.3)$ & $+109.6$ & $+107.5\;(-0.4)$ & $-19.0$ & $-23.0\;(-0.6)$ &
$+93.5$ & $+84.8\;(-3.6)$ \\
A$_\infty$B$_\infty$ &  & 
 $0.0\;(0.0)$ & & $+67.3\;(+1.6)$ & & $+53.4\;(-0.4)$ & $+576.2$ & $+84.8\;(-2.0)$ \\
&&&&&&&&\\

\underline{(401) Struct:}&&&&&&\\
A$_5$B$_1$A$_1$B$_1$ & $D0_{23}$& 
 & & & $-33.3$ & $-33.6\;(0.0)$ & &\\
&&&&&&&&\\

\underline{(601) Struct:}&&&&&&\\
A$_5$B$_1$A$_1$B$_1$ & LPS$-$3& 
 & & & $-34.1$ & & &\\
&&&&&&&&\\

\underline{Other Struct:}&&&&&&\\
A$_3$B$_1$ & $L1_2$& 
$-43.4\;(+0.4)$ & $+84.8$ & $+84.8\;(-1.4)$ & $-37.3$ & $-37.3\;(-0.1)$ &
$+77.5$ & $+77.5\;(-2.7)$ \\
A$_1$B$_3$ & $L1_2$& 
$-44.0\;(+0.3)$ & $+76.0$ & $+76.0\;(+1.8)$ & $-17.3$ & $-17.3\;(-0.8)$ &
$+78.9$ & $+78.9\;(-0.2)$ \\
A$_7$B$_1$ & $D7$ & 
$-20.8\;(+0.6)$ & $+61.9$ & $+61.9\;(-3.1)$ & $+6.8$ & $+6.8\;(-8.3)$ &
$+82.9$ & $+82.9\;(-15.8)$ \\
A$_4$B$_4$ & $D4$ & 
$-42.9\;(+1.1)$ &&&&&&\\
A$_1$B$_7$ & $D7_b$ & 
$-20.0\;(-0.1)$ & $+47.1$ & $+47.1\;(-3.3)$ & $+12.9$ & $+12.9\;(+1.9)$ &
$+56.8$ & $+56.8\;(-0.7)$ \\
A$_8$B$_1$ & Ni$_8$Nb-type&
& $+63.7$ & $+47.7\;(+0.4)$ & $+9.3$ & $-9.1\;(-4.5)$ & & \\
A$_1$B$_8$ & Ni$_8$Nb-type& 
& $+42.7$ & $+36.4\;(-1.7)$ & $+30.9$ & $+18.2\;(+13.3)$ & & \\
&&&&&&&&\\

\underline{Random:}&&&&&&\\
A$_4$B$_4$ & SQS$8_{\rm a}$& 
$-42.5\;(+0.2)$ & & & & $+12.9\;(+5.7)$ & & $+122.6\;(+1.2)$ \\
A$_4$B$_4$ & SQS$8_{\rm b}$& 
$-43.6\;(-0.2)$ & & & & $-15.2\;(-5.7)$ & & $+97.5\;(-9.7)$ \\
A$_3$B$_1$ & SQS$14_{\rm a}$&
& $+116.2$ & $+77.3\;(+7.0)$ & $+56.5$ & $+5.5\;(+7.7)$ & $+183.2$ &
$+96.8\;(+15.3)$ \\
A$_1$B$_3$ & SQS$14_{\rm b}$& 
& $+92.2$  & $+69.7\;(-7.0)$ & $+37.8$ & $-5.2\;(-7.7)$ & $+118.2$ &
$+59.8\;(-15.3)$ \\
&&&&&&&&\\
\end{tabular}
\label{tab:bigtable}
\end{table*}

The total energies of the ordered structures and end-point
constituents are obtained keeping all computational parameters
exactly equal. Specifically, we always use the same basis sets
($RK_{\rm max}=9$), charge density cutoffs ($RK_{\rm max}=19$),
muffin-tin radii $R_{\rm Au}=2.4a_0$,  $R_{\rm Ag}=R_{\rm Cu}=R_{\rm
Ni}=2.2a_0$,
maximum difference in the angular momenta in the nonspherical
Hamiltonian terms ($l_{\rm max}=4$), maximum angular momenta in
the nonspherical charge densities and potentials inside the muffin-tin
spheres ($l_{\rm max}=8$), and
equivalent {\bf k} point sets\cite{froyen89} in the evaluation of
Brillouin zone integrals. When the unit cell vectors of the ordered
compound permit, we choose a {\bf k} mesh equivalent to the 
60 special points $8 \times 8 \times 8$ fcc mesh. 
Several structures (e.g., those of $A_2B$ or $AB_2$ stoichiometry)
have reciprocal unit cell vectors which are
incommensurate with the $8 \times 8 \times 8$ mesh. In these cases
we calculate the total energies of the compounds {\it and\/} fcc
constituents with a finer {\bf k} point grid.
This procedure ensures that, due to systematic cancellation of
errors, the formation enthalpies $\Delta H(\sigma)$,
Eq.~(\ref{eq:LDAenergies}), converge much faster than the total
energies. Indeed, the tests for Cu-Au described in
Sec.~\ref{sec:GS-CuAu} show that with our choice of parameters 
$\Delta H(\sigma)$ are converged to within
2~meV/atom.

The atomic positions are relaxed using quantum mechanical
forces\cite{lapw-forces} obtained at the end of the self-consistency
iterations. Minimization of the total energy with respect to the
cell-external degrees of freedom
is done by distorting the shape of the unit cell and tracing the
decrease in the total energy. We estimate that the formation enthalpies
are converged to at least 5~meV/atom with respect to all relaxational
degrees of freedom. 

Table~\ref{tab:strucs} and its caption defines our small-unit-cell
ordered structures. Many are actually superlattices along (100),
(110), (111), (201) and (311) directions. Table~\ref{tab:bigtable}
gives the calculated LDA formation energies
[Eq.~(\ref{eq:LDAenergies})] for these Au-Ag, Cu-Au, Cu-Ag 
and Ni-Au compounds.

\subsection{The constituent strain energy}
\label{sec:Ecs}

It is well known\cite{NATO} that real-space cluster expansions with
finite-ranged interactions incorrectly predict zero formation
enthalpies per atom for coherent long-period $A_p B_q$ superlattices,
while the correct answers are non-zero and depend on the 
superlattice direction $\widehat{G}$.
The constituent strain energy term $\Delta E_{\rm CS} (\sigma)$
in Eq.~(\ref{eq:recipCE}) is specifically designed to reproduce
these superlattice energies, which are calculated directly from the
LDA as follows.

\begin{figure}
\epsfxsize=2.6in
\centerline{\epsffile{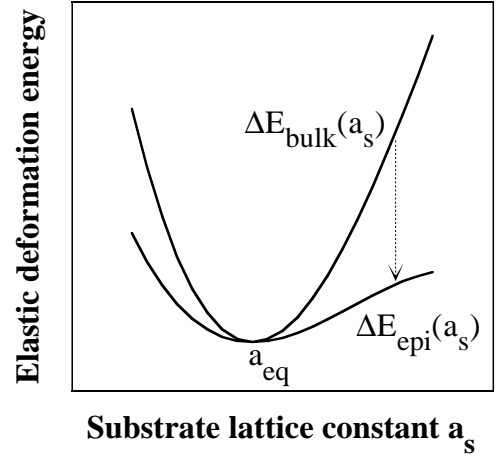}}
\vskip 5mm
\caption{A schematic illustration of the concept of the epitaxial
softening function $q(\widehat{G})$, given by the ratio of the bulk
(upper curve) and epitaxial (lower curve) deformation energies.
In the harmonic approximation $q(\widehat{G})$ is the ratio of the
curvatures of these curves at the equilibrium point.}
\label{fig:q(G)}
\end{figure}

In the long-period limit $pq \rightarrow \infty$
the interfacial energy becomes negligibly
small (${\cal O}(1/p)$) in comparison with the elastic strain energy
needed to deform the constituents to a common in-plane lattice
constant $a_s$.\cite{zwlu-SLs,paperIII} Therefore, the formation
energy per atom of  
$A_\infty B_\infty$ superlattice along $\widehat{G}$ with composition
$x$ is given by the constituent strain energy 
$\Delta E_{\rm CS}(x, \widehat{G})$, defined as the equilibrium (eq)
value of the composition-weighted sum of the energies required to
deform bulk $A$ and $B$ to the epitaxial geometry with a planar
lattice constant $a_s$:
\begin{eqnarray}
\nonumber
\Delta E^{\rm eq}_{\rm CS} (x,\widehat{G}) = \min_{a_s} \left[ x
\Delta E^{\rm epi}_{\rm A} (a_s, \widehat{G}) \right.\\
\label{eq:Ecsdef}
+ \left. (1 - x) \Delta E^{\rm epi}_{\rm B} (a_s, \widehat{G}) \right].
\end{eqnarray}
Here $\Delta E^{\rm epi} (a_s, \widehat{G})$ is the strain
energy of the material epitaxially stretched to the lattice constant
$a_s$ in the direction orthogonal to $\widehat{G}$, and then
allowed to relax along $\widehat{G}$. 
$\Delta E^{\rm epi} (a_s, \widehat{G})$ is related to the 
bulk equation of state $\Delta E^{\rm bulk}(a_s)$ via the epitaxial
softening function $q(a_s,\widehat{G})$:
\begin{equation}
\label{eq:qdef}
q(a_s, \widehat{G}) \equiv \frac{\Delta E^{\rm epi} (a_s,\widehat{G})}
{\Delta E^{\rm bulk} (a_s)},
\end{equation}
where $\Delta E^{\rm bulk}_A (a_s)$ is the energy required to
hydrostatically deform a bulk solid to the lattice constant $a_s$.
Figure~\ref{fig:q(G)} illustrates the concept of epitaxial
softening:\cite{epi-review}
when the bulk solid is deformed hydrostatically from $a_{\rm eq}$
to $a_s \neq a_{\rm eq}$, its energy rises. Energy can then be lowered
if we keep $a_x=a_y=a_s$ but relax the third lattice vector to its
equilibrium value. $q(a_s,\widehat{G})$ measures the relative energy
lowering.

\begin{figure}
\epsfxsize=2.6in
\centerline{\epsffile{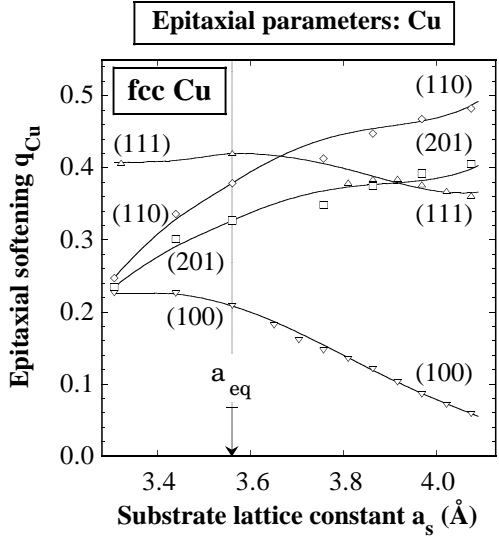}}
\vskip 5mm
\caption{$q(\widehat{G})$ of fcc Cu for principle directions as
functions of the substrate lattice parameter $a_s$. Directly
calculated LDA values are represented by open symbols, lines show the
fit using the expansion of $\gamma(\widehat{G})$ in Kubic harmonics.}
\label{fig:qCu}
\end{figure}

Figure~\ref{fig:qCu} shows the calculated LDA $q$'s for Cu,
obtained by minimizing the total energy with respect to
the lattice constant $c$ parallel to $\widehat{G}$ for each
value of the substrate lattice parameter $a_s$. As explained in
Ref.~\onlinecite{paperIII}, this treatment neglects the so-called
shear strain terms, but is exact for the high symmetry directions
(100), (111) and (110). The calculated $q_{\rm Cu} (a_s,\widehat{G})$
is seen to be a nontrivial function of the substrate lattice parameter 
$a_s$ and direction $\widehat{G}$. In contrast, the
harmonic elasticity
theory,\cite{epi-review,HB78,anastassakis,japanese} routinely used for
semiconductor systems,\cite{epi-review,japanese,wood92} gives $q$'s
which do not depend on $a_s$:
\begin{equation}
\label{eq:qharm}
q_{\rm harm} (\widehat{G}) = 1 - \frac{B}{C_{11} +
\Delta \; \gamma_{\rm harm} (\widehat{G})},
\end{equation}
where $\gamma_{\rm harm} (\widehat{G})$ is a geometric function
of the spherical angles formed by $\widehat{G}$:
\begin{eqnarray}
\nonumber
\gamma_{\rm harm} (\phi, \theta) = \sin^2 (2\theta) + \sin^4 (\theta)
\sin^2 \\
\label{eq:gamma_harm}
= \frac{4}{5} \sqrt{4\pi} [K_0(\phi,\theta) - \frac{2}{\sqrt{21}}
K_4(\phi,\theta) ],
\end{eqnarray}
and $K_l$ are the Kubic harmonics of angular momentum $l$.
Figure~\ref{fig:qCu} shows that the harmonic approximation manifestly
breaks down for large epitaxial strains in metals since there are
several important
{\it qualitative\/} differences between the behavior in
Fig.~\ref{fig:qCu} and that predicted by the harmonic elasticity.
First, $q(a_s,\widehat{G})$ strongly depends on the substrate
lattice constant, while the harmonic $q_{\rm harm} (\widehat{G})$
does not. Second, the harmonic expression gives a definite
order of $q(\widehat{G})$ as a function of the direction, i.e., either
(100) is the softest and then (111) {\it must\/} be the hardest, or
vice versa. This order does not hold for large deformations.
For instance, (201) becomes the softest direction for $a_s \ll a_0$
and (110) is the hardest for $a_s \gg a_0$ in Cu. Finally,
$q(100)$ exhibits a particularly dramatic softening for $a_s \gg a_0$,
which has important consequences for the constituent strain energy
and stability of superlattices along this direction.\cite{paperIII}

The above mentioned properties of $q_{\rm Cu}$ can be described
by generalizing Eq.~(\ref{eq:gamma_harm}) for
$\gamma$ to higher Kubic harmonics and strain-dependent expansion
coefficients:
\begin{equation}
\label{eq:gamma_general}
\gamma(a_s,\widehat{G}) = \sum_{l=0}^{l_{\rm max}}
b_l(a_s) \, K_l (\widehat{G}),
\end{equation}
which has the property that in the harmonic limit ($a_s
\rightarrow a_0$) all expansion coefficients with angular momenta higher
than 4 tend to zero, reproducing $\gamma_{\rm harm}$ from
Eq.~(\ref{eq:gamma_harm}). Due to the cubic symmetry, only terms with
$l=0,4,6,8,10,12, \ldots $ enter in this expansion.
Detailed discussion of the nonlinear epitaxial strain properties
of elemental metals will be given in a separate
publication.\cite{paperIII}

\begin{figure}
\epsfxsize=2.6in
\centerline{\epsffile{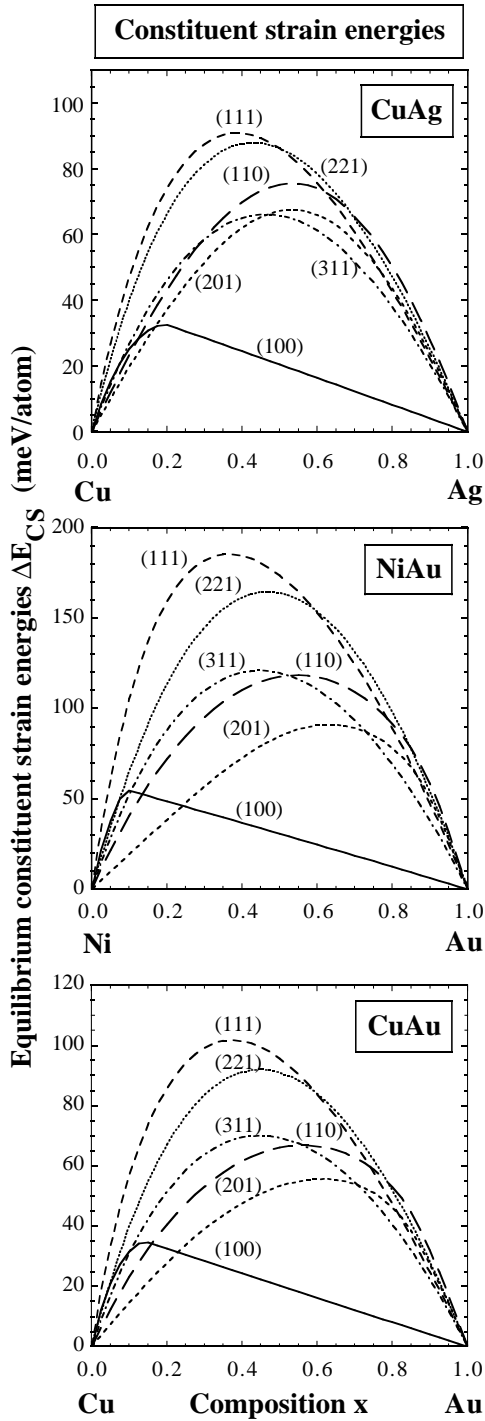}}
\vskip 5mm
\caption{Equilibrium constituent strain energies for Cu-Au, Ni-Au and
Cu-Ag. The constituent strain energy of Ag-Au is negligibly small and
therefore not shown.}
\label{fig:ECS}
\end{figure}

The constituent strain energy
$\Delta E_{\rm CS}^{\rm eq} (x,\widehat{G})$
is calculated numerically from Eq.~(\ref{eq:Ecsdef}) 
using the direct LDA values of $\Delta E^{\rm epi} (a_s,\widehat{G})$
for six principle directions. The obtained 
$\Delta E_{\rm CS}^{\rm eq}$ for these directions
are shown in Fig.~\ref{fig:ECS}, illustrating several properties of
the constituent strain which cannot be reproduced by the harmonic
theory.\cite{dblaks92} First, the curves in Fig.~\ref{fig:ECS} are
skewed to different sides, while the harmonic $\Delta E_{\rm CS}^{\rm
eq}$ must be all skewed to the same side. Second, the calculated
$\Delta E_{\rm CS}^{\rm eq}$ cross for different directions, a
property not allowed by the harmonic functional form.
These crossings lead to (201) as the softest
direction below $x \approx 0.2$, and (110) as the hardest
for Au-rich superlattices, while the harmonic theory
gives $\Delta E_{\rm CS}^{\rm eq}(111)$ as the highest and 
$\Delta E_{\rm CS}^{\rm eq}(100)$ as the lowest 
constituent strain for all compositions of the 
studied noble metal alloys. The behavior of
$\Delta E_{\rm CS}^{\rm eq}$ for (100) is particularly
interesting, since the curves in Fig.~\ref{fig:ECS} abruptly
change slope around $x \approx 0.15$ and have very low values
for $x > \frac{1}{4}$. As we show in Ref.~\onlinecite{paperIII},
this is a manifestation of the low energy cost of deforming fcc Cu
into the body-centered tetragonal structure along the epitaxial Baines
path. Small constituent strain of (100) superlattices has profound
influence on the predicted ground states of Cu-Au (see
Sec.~\ref{sec:GS-CuAu}).

The constituent strain energy for arbitrary direction $\widehat{G}$
is then obtained by interpolating between the principle directions
using the following expansion in Kubic harmonics:
\begin{equation}
\label{eq:Ecs_expansion}
\Delta E_{\rm CS} (x,\widehat{G}) = \sum_{l=0}^{l_{\rm max}}
c_l(x) \, K_l (\widehat{G}).
\end{equation}
We have taken $l_{\rm max} = 10$, which gives five
composition-dependent fitting coefficients determined from a fit to
the directly calculated values [Eq.~(\ref{eq:Ecsdef})] for six principle
directions. The characteristic errors of
this fit at the equiatomic composition are $1-2$~meV/atom.
Equation~(\ref{eq:Ecs_expansion}) is then used in
Eqs.~(\ref{eq:Ecs(sigma)def})--(\ref{eq:Jcsdef}).

\subsection{Constructing the Cluster Expansion}
\label{sec:fitCE}

\begin{figure*}
\epsfxsize=5.2in
\centerline{\epsffile{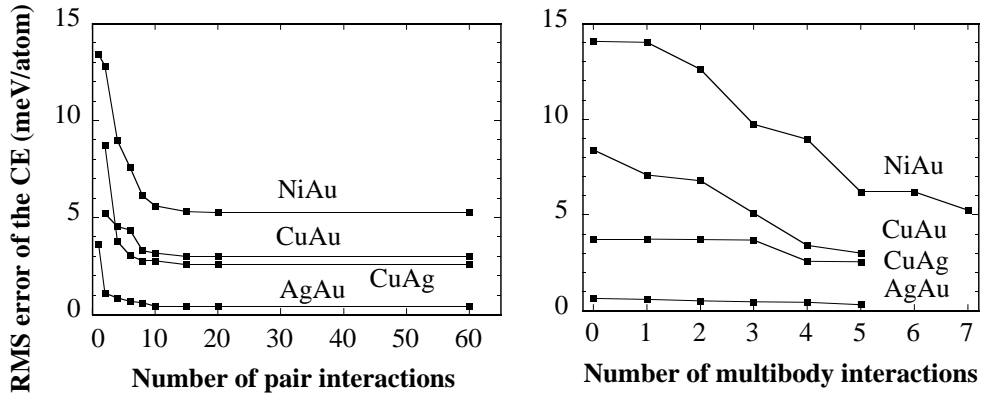}}
\vskip 5mm
\caption{Root-mean-square errors $\Delta_{\rm rms}$ of the cluster
expansions for Ag-Au, Cu-Ag, Cu-Au and Ni-Au as functions of the
number of pair and multibody interactions.}
\label{fig:CE-RMS}
\end{figure*}

Once we have a closed-form expresion for the equilibrium constituent
strain energy $\Delta E_{\rm CS} (\sigma)$ and a set $\{\Delta
H^{\rm LDA} (\sigma) \}$ of $T=0$ formation enthalpies, we determine
the unknown cluster interactions of Eq.~(\ref{eq:recipCE}) in the
following two-step process:

{\it First,\/} the total energies of all structures from
Table~\ref{tab:bigtable} are used in the fit to investigate the
behavior of the root-mean-square (rms) error $\Delta_{\rm rms}$ of the
fit, Eq.~(\ref{eq:CErms}), as a function of the number of real-space pair
and multibody interactions. Reciprocal space CE allows to add pair
interactions systematically in the order of increasing intersite
separation, up to any number of near-neighbor shells. 
The {\bf k}-space smoothness criterion in
Eq.~(\ref{eq:CErms}) automatically selects optimally short-ranged
interactions and chooses physically important pair interactions
which are essential to produce a good fit to the directly calculated
LDA energies. The dependence of the rms error on the number of pair
and multibody interactions is shown in Fig.~\ref{fig:CE-RMS}.
Figure~\ref{fig:CE-RMS}(a) is obtained by fixing the
number of multibody interactions, and varying the number of pair
interactions. It shows that in all systems the cluster
expansion is well converged using 10 to 20 pair interactions.
The convergence rate is fastest for Ag-Au and slowest for Ni-Au,
which we attribute to increasing size mismatch going from Ag-Au
to Ni-Au, with Cu-Ag and Cu-Au exhibiting intermediate convergence
rates.

Selection of important multibody interactions is more delicate.
The number of pair interactions is fixed to a converged value
(20 or more), and a large set of 3- to 4-body figures is tested as to 
whether it improves the rms
error of the overall fit. It is retained in the CE only if
$\Delta_{\rm rms}$ decreases considerably. During the fitting process,
we also monitor the overall stability of the CE, as measured by a
change in other multibody interactions upon the addition of a
particular figure. Unstable behavior usually signals of linear
dependencies in the chosen set of clusters and an ill-conditioned
inverse problem, necessitating a different
choice of $\{J_f\}$. Figure~\ref{fig:CE-RMS}(b) shows the convergence
of the CE with respect to the number of multibody interactions,
keeping $N_{\rm pairs}$ equal to their converged values. An important
thing to notice is that the multibody interactions produce a
decrease in the rms error which is of the same magnitude as that due
to the pair interactions. Furthermore, the effect of multibody
interactions is largest in Ni-Au, and decreases in order of decreasing
size mismatch, becoming negligible in Ag-Au.

{\it In the second step\/} we test the stability of the fit and its
predictive power. Using the trial set of figures obtained in the
previous step, we exclude several structures which are fit rather
well (e.g., $Z2$, $\beta 2$, and $L1_2$ in Ni-Au), and repeat the
fit, obtaining new values of the effective cluster interactions.
These values are used to predict the total energies of the structures
excluded from the fit. If the change in $\Delta H_{\rm CE} (\sigma)$
is not acceptable (more than few meV/atom), we return
to the first step to search for a better set of interactions. 
The most severe test is to exclude structures with the poorest fit to
their formation enthalpies, e.g., $SQS14_a$ and $SQS14_b$ in Ni-Au.
If the predicted formation energy does not change significantly, the
chosen set of figures is considered to be stable and predictive.
The final cluster expansion is produced by using this set of figures
and all structures from Table~\ref{tab:bigtable}.

\begin{figure}
\epsfxsize=2.7in
\centerline{\epsffile{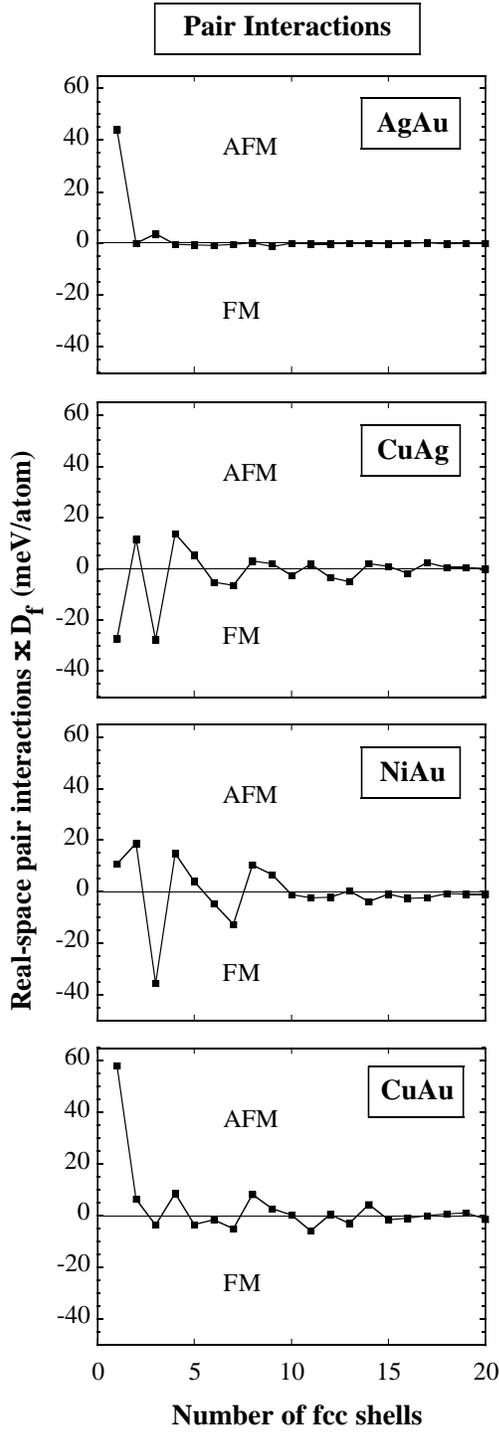}}
\vskip 5mm
\caption{Real space pair interactions for the studied noble metal
alloy systems.}
\label{fig:pair-ECIs}
\end{figure}

Figure~\ref{fig:pair-ECIs} shows the calculated pair interactions as
function of the near-neighbor fcc shell. There are several
noteworthy trends in the four alloy systems:

(i) Only in Ag-Au and Cu-Au are the nearest-neighbor pair
interactions dominant: in Cu-Ag the 1-st and 3-rd neighbor
pair interactions are of similar magnitude, while the 3-rd neighbor
interaction dominates in Ni-Au.

(ii) The dominant interactions have signs consistent with the
observed phase diagrams: Ag-Au and Cu-Au have positive
(``antiferromagnetic'') nearest-neighbor pair 
interactions $J_2$, corresponding to the tendency towards complete
miscibility and ordering at low temperatures. The behavior of Ni-Au,
in spite of {\it positive\/} 1-st and 2-nd neighbor pair interactions,
is dominated by the ``ferromagnetic'' 3-rd neighbor interaction $L_2$
(which causes phase separation at low temperatures). Both dominant 1-st
and 3-rd neighbor pair interactions in Cu-Ag are negative, implying a
miscibility gap. The constituent strain energy 
$\Delta E_{\rm CS}^{\rm eq}$ is always positive and therefore
increases the propensity for incoherent phase separation.

(iii) Although the nearest-neighbor pair interaction is clearly
dominant in Cu-Au, other pair interactions show a long-ranged
oscillatory behavior extending over approximately 15 shells. As found
in other systems,\cite{dblaks92,italians} this is a direct consequence
of the atomic relaxation caused by the constituent size mismatch
between Cu and Au. The pair interactions are slowly decaying in Cu-Ag
and Ni-Au, too.

\begin{figure}
\epsfxsize=2.7in
\centerline{\epsffile{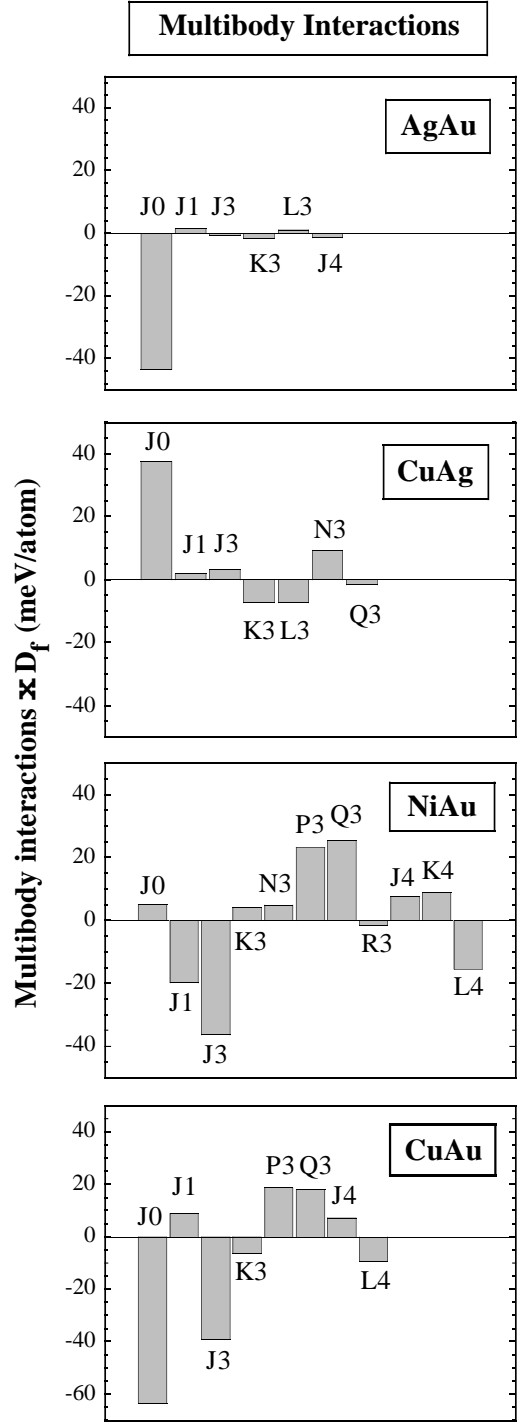}}
\vskip 5mm
\caption{Multibody interactions for the studied noble metal
alloy systems.}
\label{fig:MB-ECIs}
\end{figure}

The calculated multibody interaction energies are shown in
Figure~\ref{fig:MB-ECIs}. $J_1$ is the point interaction, $J_3$, $K_3$,
$N_3$, ..., are triplets and $J_4$, $K_4$, and $L_4$ are four-point
clusters in increasing order of interatomic separation
(see Lu {\it et al.\/}\cite{zwlu-agau} for a full description of
the clusters). Figure~\ref{fig:MB-ECIs} illustrates the importance of
the multibody terms in our Hamiltonian.

\subsection{Finding the $T=0$ ground states and $T>0$ properties}
\label{sec:simann}

Having parametrized the configurational energies in terms of the
mixed-space cluster expansion Eq.~(\ref{eq:recipCE}), we can
use it with established statistical methods to predict
various structural properties: $T=0$ ground states, order-disorder
transition temperatures, configurational entropies, free
energies, phase stabilities and atomic short-range order
parameters.
Due to the presence of both reciprocal and real space terms in
the Hamiltonian~(\ref{eq:recipCE}), traditional
techniques, e.g., the Cluster Variation Method, are not
readily applicable. Monte Carlo simulations must be used instead to
calculate statistical properties at finite temperatures. The
basic computational algorithm is as follows. We adopt the
Metropolis algorithm in the canonical ensemble (fixed
composition). For each attempted spin flip, the change in the multiplet
interaction energy is 
evaluated in the real space. To obtain the reciprocal space energy
(constituent strain and pair interaction energies), the Fourier
transform of the spin function $S({\bf R}_i, \sigma)$ is needed. It
can be calculated either with the help of the Fast Fourier Transform
(FFT) or evaluated directly taking advantage of the special method
described in Ref.~\onlinecite{MCpaper}, which is much more economical:
if the total number of sites in the simulation box is $N$,
a full FFT has to be done only once after approximately every
$\sqrt{N}$ accepted spin flips, which makes the whole computational
effort for this special method scale as $N^{1.5}$.

A simulation box of $N=4096$ atoms ($16 \times 16 \times 16$) is used
to calculate all thermodynamic properties presented in this paper.
The transition temperatures are computed by cooling the system
from high temperatures and monitoring the discontinuities
in the average energy and peaks in heat capacity. To eliminate possible
hysteresis effects, the resulting low-temperature configurations are
gradually heated up past the transition point. The former process
provides the lower bound on the transition temperature, $T_1$, while
the latter gives the upper bound, $T_2$. The heating and
cooling rates are such that $T_1$ and $T_2$ differ by no more than 20K,
an insignificant uncertainty compared to the inaccuracies
of the LDA calculations and the fit errors of the cluster expansion.
1000 flips/site and a temperature decrease of 2\% for each Monte Carlo
step are usually sufficient, although in a few cases the results are
checked using 2000 flips/site and $0.5\%$ temperature change.

Zero temperature ground states are found by cooling the system to $T=0$
and checking whether the energy of the final configuration lies on the
convex hull. This process is repeated for several random number
seeds and starting temperatures, always yielding configurations with
similar (usually identical) energies. We explore many equally spaced
compositions with an interval $\Delta x = 0.05$. The number of
possible configurations for each $x$ is
$N_{\rm conf} = \frac{N!}{(xN)!(N(1-x))!}$.

Configurational entropy of the disordered alloys at finite
$T$ is computed from the energy {\it vs.\/} temperature
curves obtained by cooling the system from
very high (``$T=\infty$'') temperatures. The following thermodynamic
formula gives the configurational entropy at temperature $T$:
\begin{equation}
\label{eq:Sconf}
\Delta S_{\rm conf} (T) = \Delta S_{\rm ideal} + E(T)/T - k_B \int_0^\beta
E(\beta') \, d\beta',
\end{equation}
where $\beta = 1/k_BT$ and $\Delta S_{\rm ideal} = k_B [ x \log x + (1-x)
\log (1-x) ]$ is the configurational entropy of an ideal solid solution.

\section{Results}
\label{sec:results}

\subsection{$T=0$ Ground States}
\label{sec:GS}

\subsubsection{Ground states of Cu-Au}
\label{sec:GS-CuAu}

\begin{figure}[t]
\epsfxsize=3.0in
\centerline{\epsffile{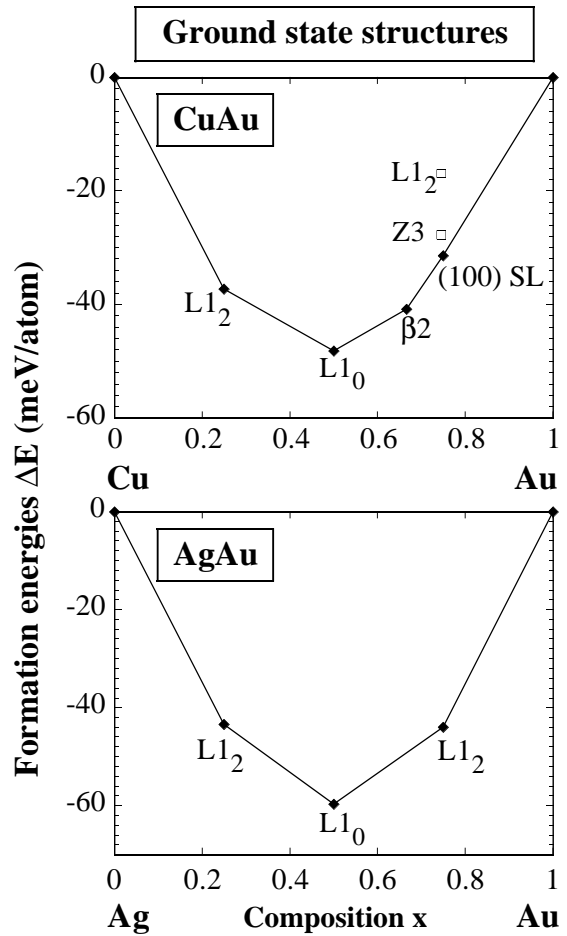}}
\vskip 5mm
\caption{$T=0$~K ground state lines for Cu-Au and Ag-Au obtained from
simulated annealing calculations. $L1_2$ CuAu$_3$ is not only above
the ground state line, but also has a higher formation enthalpy than
other structures at the same composition, e.g. LDA calculations place
the formation enthalpy of $Z3$ below that of $L1_2$. Plots for
Cu-Ag and Ni-Au are not shown since these systems phase separate at
$T=0$~K.}
\label{fig:GSL}
\end{figure}

Figure~\ref{fig:GSL} shows the calculated $T=0$ ground state lines of
Cu-Au and Ag-Au which were obtained from simulated annealing
quenches of a $16 \times 16 \times 16$ system. In Cu-Au, we 
find the $L1_2$ (Cu$_3$Au) and $L1_0$ (CuAu) structures as the
stable ground states of Cu-rich alloys, in agreement with the
existing phase diagram
data.\cite{cuau_phase_D,hultgren,hansen,massalski}
These data also list $L1_2$ as the stable low-temperature phase of
CuAu$_3$. However, we find new, previously unsuspected ground states
of Au-rich compounds, all belonging
to the family of (001) superlattices. At $x=\frac{2}{3}$ we find
a stable $\beta 2$ (CuAu$_2$) phase (prototype MoSi$_2$), which is a 
Cu$_1$Au$_2$ superlattice along (001). At $x=\frac{3}{4}$, our
cluster expansion predicts that a complex
Cu$_1$Au$_4$Cu$_1$Au$_4$Cu$_1$Au$_2$Cu$_1$Au$_2$ (001) 
superlattice falls on the convex hull, although its energy is
less than 2~meV below the tieline connecting $\beta 2$ (CuAu$_2$) and
Au. Furthermore, even the directly calculated LDA
enthalpy of formation of $Z3$  (which is a Cu$_1$Au$_3$ (001)
superlattice) is considerably lower than that of $L1_2$ CuAu$_3$.

We carefully checked whether the predicted new LDA ground states 
for Au-rich Cu-Au alloys artifacts of some approximation in our LDA
calculations or the fit error of the cluster expansion.  The latter
possibility was
quickly dismissed, since the directly calculated LDA enthalpies of
formation for $L1_0$, $\beta 2$, $L1_2$ and $Z3$ agreed with the values
derived from the cluster expansion to better than 2~meV/atom (see
Table~\ref{tab:bigtable}), while the new (100) SL ground state is
14~meV/atom below $L1_2$. To address the former possibility, we
performed careful convergence tests for  $L1_0$, $\beta 2$, $L1_2$ and
$Z3$ with respect to the 
plane wave cutoff and number of {\bf k} points in the first Brillouin
zone. The cutoff was increased from $RK_{\rm max}=9$ to $RK_{\rm
max}=11$ and the density of the Brillouin zone mesh 
was doubled from $8 \times 8 \times 8$ to $16 \times 16 \times 16$, 
an eightfold increase in the total number of {\bf
k} points. These tests showed that the formation enthalpies of
$L1_0$, $\beta 2$ and $L1_2$ were converged to within 1 meV/atom
with respect to the size of the basis set and the number of {\bf k}
points. Further, we checked how the choice of muffin-tin radii affected
$\Delta H$. Varying $R_{MT} ({\rm Au})$ between $2.3a_0$ and $2.5a_0$
changed the formation enthalpies by at most 2~meV/atom and did not
shift the relative stabilities of phases. Finally, we repeated these
calculations using the Perdew-Zunger\cite{perdewzunger}
parametrization of the Ceperley-Alder\cite{ceperleyalder} LDA
functional, as well as the generalized gradient approximation (GGA) of
Perdew and Wang,\cite{ggapw91} and found insignificant (about
2~meV/atom) changes in the formation enthalpies.
Inclusion of the spin-orbit interaction in the second variation
procedure\cite{spin-orbit} changed the formation enthalpy of $L1_0$
(CuAu) by only $3.7$~meV/atom (from $-48.2$ to $-51.9$), 
indicating that it is not
important for the energetics of Cu-Au. This conclusion is in line with
the findings of Ref.~\onlinecite{wei88} that the spin-orbit
interaction influences the band structure but has little effect on
equilibrium lattice properties.
Therefore, we conclude that {\it state-of-the-art 
first-principles density functional calculations do
not predict $L1_2$ to be a stable $T=0$ ground state of CuAu$_3$.\/}
It is possible that van der Waals interactions, omitted by the LDA and
important for large, polarizable atoms such as Au, can affect the
formation energies and hence the ground states of Cu-Au.

We next analyze the possibility that the correct $T=0$
ground state around $x=\frac{3}{4}$ is not $L1_2$ as has been
assumed in the literature before.
Although most
compilations\cite{cuau_phase_D,hultgren,hansen,massalski} of
binary alloy phase diagrams give $L1_2$ as the stable structure of
CuAu$_3$, the experimental evidence\cite{ogawa51,batterman57,kubiak}
seems inconclusive because of the difficulties in obtaining
equilibrated long-range ordered samples. X-ray
studies\cite{batterman57} have found superlattice peaks consistent
with the cubic $L1_2$ structure, but only very broad low-order
reflections have been observed. These superlattice lines could not be
sharpened by any heat treatment.\cite{batterman57} It is not clear to
us if the X-ray reflections can be reindexed according to some other
non-$L1_2$ phase. It is also possible that at elevated ($T \approx
500$~K)  temperatures $L1_2$ is stabilized by the entropy
(configurational {\it and\/} vibrational), while another transformation
to the low-energy structure should occur but is kinetically inhibited 
below $500$~K. The biggest experimental obstacles to
verifying our predictions seem to be low diffusion rates below
the ordering temperature of CuAu$_3$, $T_c \approx 500$~K.

Next we discuss the experimental signatures of the new LDA ground
state 
structures. MoSi$_2$-type $\beta 2$ CuAu$_2$ has a superlattice
reflection at $(\frac{2}{3}00)$, but CuAu$_3$ (100) superlattice has
reflections at (100) and
$(\frac{1}{3}00)$. These reflections also manifest
themselves in the predicted atomic short-range order of the disordered
alloys (for details see Ref.~\onlinecite{paperII}).

\subsubsection{Ground states of Ag-Au, Cu-Ag and Ni-Au}
\label{sec:GS-others}

The ground state line of Ag-Au is shown in Figure~\ref{fig:GSL}(b),
exhibiting $L1_2$ (Ag$_3$Au), $L1_0$ (AgAu) and $L1_2$ (AgAu$_3$)
stable low-temperature phases. Experimentally, these alloys are known
to be completely miscible,\cite{hultgren,hansen,massalski} and there
are several 
indications\cite{AgAu-lowT} that they would order below 200 K if not
for the very low diffusion rates. Theoretical transition
temperatures and short-range order patterns, as well as a complete
discussion are given by Lu and Zunger.\cite{zwlu-agau}

The calculated ground states of Cu-Ag and Ni-Au are found to be phase
separation, in agreement with the 
experimental enthalpy data.\cite{hultgren} Neither alloy has a single
ordered or disordered structure with negative enthalpy of formation
and therefore there are 
no stable $T=0$ ground states except the phase-separated alloy.

\subsection{Mixing enthalpies}
\label{sec:Hmix}

\begin{table*}
\squeezetable
\caption{Calculated mixing enthalpies of disordered Cu$_{1-x}$Au$_x$
alloys compared with the values obtained by other studies and
experimental measurements (in meV/atom). FLAPW is the full-potential
linearized augmented plane wave method, LMTO -- linearized
muffin-tin-orbitals method, KKR -- Korringa-Kohn-Rostoker
multiple scattering method, ASA -- atomic-sphere approximation, CPA
-- coherent potential approximation, CWM - Connolly-Williams cluster
expansion, MSCE -- mixed-space cluster expansion used in this
study, ``Rel.'' -- incorporating atomic relaxations, and ``Unrel.''
-- neglecting atomic relaxations.}
\begin{tabular}{l|ccccccc}
Composition & Expt.\tablenotemark[6] & This & Wei & 
Amador & Terakura & Ruban & Weinberger \\
       &  & study & {\it et al.\/}\tablenotemark[1] & 
{\it et al.\/}\tablenotemark[2] & {\it et al.\/}\tablenotemark[3] 
& {\it et al.\/}\tablenotemark[4] & {\it et al.\/}\tablenotemark[5]\\
  &  & FLAPW & FLAPW & LMTO-ASA & ASW & LMTO-ASA & KKR-ASA\\
 & & MSCE & CWM & CWM & CWM  & CPA & CPA \\
 & & (Rel.) & (Rel.) & (Unrel.) & (Unrel.) & (Unrel.) & (Unrel.)\\
\tableline
$\underline{\Delta H_{\rm mix} (T=\infty)}$ &&&&&&&\\
Cu$_{0.75}$Au$_{0.25}$ & & $+2.6$ & $+46.3$ & $+59$ & $+26.9$ &
$+54.6$ & $-27$ \\
Cu$_{0.50}$Au$_{0.50}$     & & $+1.6$ & $+38.0$ & $+61$ & $+30.4$ &
$+44.3$ & $-57$ \\
Cu$_{0.25}$Au$_{0.75}$ & & $+5.4$ & $+18.6$ & $+39$ & $+20.4$ &
$+19.8$ & $-31$ \\
\\
$\underline{\Delta H_{\rm mix} (T=800\;{\rm K})}$&&&&&&& \\
Cu$_{0.75}$Au$_{0.25}$ & $-46$\tablenotemark[7] & $-17.3$ &         & $-6$ \\
Cu$_{0.50}$Au$_{0.50}$ & $-53$\tablenotemark[7] & $-19.3$ & $-16.9$ & $-5$ \\
Cu$_{0.25}$Au$_{0.75}$ & $-31$\tablenotemark[7] &  $-1.2$ & $-2.6$  & $+8$ \\
\\
$\underline{\Delta H_{\rm mix} (T=0\;{\rm K})}$&&&&&&& \\
$L1_2$ Cu$_3$Au & $-74$ & $-37.3$ & $-36.0$ & & $-65.0$ & $-60.7$ &
$-54$ \\
$L1_0$ CuAu     & $-91$ & $-48.2$ & $-62.9$ & & $-69.7$ & $-83.4$ & $-76$ \\
$L1_2$ CuAu$_3$ & $-59$ & $-17.3$ & $-26.4$ & & $-34.0$ & $-56.1$ & $-47$ \\
\end{tabular}
\tablenotetext[1]{Ref.~\protect\onlinecite{wei87} using the
Connolly-Williams structures (relaxation of $L1_0$ only).}
\tablenotetext[2]{Ref.~\protect\onlinecite{amador}.}
\tablenotetext[3]{Ref.~\protect\onlinecite{terakura}.}
\tablenotetext[4]{Ref.~\protect\onlinecite{ruban95}.}
\tablenotetext[5]{Ref.~\protect\onlinecite{weinb94}.}
\tablenotetext[6]{Ref.~\protect\onlinecite{hultgren}.}
\tablenotetext[7]{Values obtained at $T=720$K.}
\label{tab:Hmix}
\end{table*}

It is interesting to compare the calculated mixing enthalpies of
disordered Cu-Au alloys with the available theoretical and experimental
data. Table~\ref{tab:Hmix} summarizes the values of $\Delta H_{\rm
mix} (x,T)$ for the completely random ($T=\infty$), short-range
ordered ($T=800$~K) and completely ordered ($T=0$~K) Cu-Au alloys at
compositions $x=\frac{1}{4}$, $\frac{1}{2}$ and $\frac{3}{4}$. 
Several important points are apparent from this table:

(i) Studies\cite{amador,terakura,ruban95} which have completely 
neglected atomic
relaxations predict a substantially positive enthalpy of formation for
the completely random alloy. In our calculations, relaxations in the
random alloy reduce $\Delta H_{\rm mix} (T=\infty)$ by a large amount,
bringing it down to essentially zero.

(ii) Comparison of the present results for the $T=\infty$ random alloys
with those of Wei {\it et al.\/}\cite{wei87} shows the influence of
the number of 
structures included in the cluster expansion. Since Wei {\it et al.\/}
used the same FLAPW method\cite{lapw}, but included a set of only five
high-symmetry ordered 
structures [$A1$ (Cu), $L1_2$ (Cu$_3$Au), $L1_0$ (CuAu), $L1_2$
(CuAu$_3$) and $A1$ (Au)], the atomic relaxation effects were included
incompletely.
Indeed, their treatment gives much larger mixing enthalpies of the
random Cu-Au alloys than the present work employing approximately
30 low-symmetry structures with large relaxations. Therefore we
conclude that the Connolly-Williams set of five ordered structures
cannot correctly capture the large decrease of the mixing
enthalpy of random Cu-Au alloys caused by the atomic relaxations.

(iii) The good agreement between the relaxed (this study) and 
``unrelaxed'' (Wei {\it et al.\/}\cite{wei87}) values of
$\Delta H_{\rm mix}$ at $T=800$~K suggests that
the short-range order in Cu-Au tends to decrease the role of the atomic
relaxations. This effect can be qualitatively explained on the basis
of the ordering tendency towards high-symmetry structures which have
little or no relaxation energy ($L1_2$ and $L1_0$ in Cu-rich alloys).

(iv) The mixing enthalpies of the random alloy calculated by
Weinberger {\it et al.\/}\cite{weinb94} using the coherent-potential
approximation (CPA) differ strongly not only
from those obtained using the cluster expansion
methods,\cite{wei87,amador,terakura} but also 
from the numbers given in the CPA work of Ruban, Abrikosov, and
Skriver.\cite{ruban95} Since the CPA of Weinberger {\it et
al.\/}~\cite{weinb94} neglects the (a) atomic relaxation, (b) charge
transfer and (c) short-range order, which all lower the
formation energies, the negative values obtained
by Weinberger {\it et al.\/}\cite{weinb94} are very puzzling.

(v) There are significant discrepancies between the best
calculated and experimentally measured\cite{orr60/1,oriani57,hultgren}
values of  $\Delta H_{\rm mix}$ at both $T=0$~K and $T=800$~K.
At present these discrepancies are hard to explain since the
available general potential LDA
calculations\cite{wei87,lu-big,LASTO88} of $\Delta H(L1_2)$ and
$\Delta H(L1_0)$ agree with each other reasonably well.
On the other hand, formation energies in Cu-Au are numerically very
small and present a severe test for any first-principles model
of electronic exchange-correlation. It is noteworthy that several
less accurate first-principles calculations, using the atomic-sphere
approximation (ASA), have achieved better agreement with the
experimental enthalpies of formation than the state of the art
general potential techniques. We consider this to be fortuitous.
In all cases, LDA calculations correctly predict the relative
magnitudes of $\Delta H$ for $L1_2$ and $L1_0$, as well as reproduce
measured asymmetry in formation enthalpies towards more
negative values of $\Delta H_{\rm mix}$ for Cu-rich alloys.

\subsection{Order-disorder transition temperatures}
\label{sec:Tcs}

Order-disorder transitions have been investigated at
compositions ($x=\frac{1}{4}$, $\frac{1}{2}$, $\frac{2}{3}$ and
$\frac{3}{4}$) using 
the Monte Carlo simulation technique described in
Sec.~\ref{sec:simann}. The resulting transition temperatures, $T_c$,
are given in Table~\ref{tab:Tc}.
All transitions are found to be first order, involving
discontinuities in the energy and correlation functions.
At $x=\frac{1}{4}$ we find a transition from the disordered state
to long-range ordered $L1_2$ Cu$_3$Au at $T_c=530$~K, which is
only $130$~K lower than the experimentally observed transition
temperature. For the equiatomic alloy at $x=\frac{1}{2}$ the
calculated and experimental transition temperatures agree
to a few degrees Kelvin. However, we do not find
the CuAu II phase which exists in a narrow temperature
range between 658~K and 683~K. This phase is stabilized
by the free energy differences between $L1_0$ and long-period
superstructures of $L1_0$ which are as small
as 1~meV/atom\cite{paxton} and therefore beyond the accuracy of
self-consistent LDA calculations.

\begin{table}
\caption{Calculated order-disorder transition temperatures (in K) for
Cu-Au. $A1$ denotes the configurationally disordered fcc phase, and
n/a means that the transition has not been observed (either
experimentally or in the Monte Carlo simulation).}
\begin{tabular}{cccc}
Compo- & Tran- & Expt. & This \\
sition & sition &      & study \\
\tableline
Cu$_{0.75}$Au$_{0.25}$ & $A1 \rightarrow L1_2$  & 663 & 530 \\
Cu$_{0.50}$Au$_{0.50}$ & $A1 \rightarrow L1_0$  & 683/658\tablenotemark[1]& 660 \\
Cu$_{0.33}$Au$_{0.66}$ & $A1 \rightarrow \beta 2$ & n/a  & 735 \\
Cu$_{0.25}$Au$_{0.75}$ & $A1 \rightarrow L1_2$  & $\approx 500$ & n/a \\
                & $A1 \rightarrow \beta 2 + A1$ & n/a  & 750 \\
                & $\beta 2 + A1 \rightarrow (100) {\rm SL}$ & n/a & 680 \\
\end{tabular}
\tablenotetext[1]{CuAu undergoes a transition to CuAu-II at 683 K,
subsequently transforming into $L1_0$ CuAu-I at 658 K.}
\label{tab:Tc}
\end{table}

\begin{table*}
\caption{The experimentally measured\protect\cite{hultgren} entropy of
formation $\Delta S_{\rm tot}^{\rm form}$, the calculated
configurationl entropy  $\Delta S_{\rm conf}^{\rm calc}$ and the
derived non-configurational entropy of formation, $\Delta S_{\rm
non-conf}^{\rm form}$. All values are given in units of
$k_B$/atom.}
\begin{tabular}{cdddddd}
System & x & $T$~(K) & $ \Delta S^{\rm form}_{\rm tot}$ 
&  $\Delta S_{\rm ideal}$
&  $\Delta S_{\rm conf}^{\rm calc}$
&  $\Delta S_{\rm non-conf}^{\rm form} = $\\
 & & & & & & $\Delta S^{\rm form}_{\rm tot} - 
       \Delta S_{\rm conf}^{\rm calc}$ \\
\tableline
Cu-Au & 0.5   &  800 & 0.73 & 0.69 & 0.57 & 0.16 \\
Ag-Au & 0.5   &  800 & 0.52 & 0.69 & 0.62 & $-$0.10 \\
Cu-Ag & 0.141 & 1052 & 0.77 & 0.41 & 0.40\tablenotemark[1] & 0.37 \\
Ni-Au & 0.5   & 1100 & 1.04 & 0.69 & 0.56 & 0.48 \\
\end{tabular}
\tablenotetext[1]{This value was obtained at $T=1136$~K, since a
coherent phase separation starts at lower temperatures.}
\label{tab:Snon-conf}
\end{table*}

For $x=\frac{3}{4}$ we obtain a sequence of transformations, the first
one occuring at $T=750$~K from the disordered $A1$ phase to a coherent
two-phase mixture of $\beta 2$ and $A1$. Then a subsequent transition at
$T=635$~K takes CuAu$_3$ into the long-range ordered (100) superlattice
which is predicted to be the stable $T=0$ ground state at that
composition (see Sec.~\ref{sec:GS-CuAu}). The calculated transition at
$x=\frac{2}{3}$ goes straight into the $\beta 2$ phase at $T=735$~K.
Therefore, a two-phase $\beta 2 + A1$ field is predicted to exist at
temperatures somewhere between 635~K and 730~K and around
$x=\frac{3}{4}$. These predictions reflect the LDA. As stated in
Sec.~\ref{sec:GS-CuAu}, corrections to the LDA might be significant.

\subsection{Non-configurational entropy}
\label{sec:Snc}

The effect of the non-configurational entropy (electronic,
vibrational, etc.) on the alloy phase stability has recently attracted
considerable
interest.\cite{chris-Ni3V,watson84,fultz-CuAu,fultz-NiAl/1,fultz-NiAl/2,fultz-FeAl,fultz-FeCr,fultz-NiV,gerardo-svib,althoff-NiAl,desplat96}
For instance, it has been
suggested\cite{fultz-CuAu,fultz-NiAl/1,fultz-NiAl/2,fultz-FeAl,fultz-FeCr,fultz-NiV,gerardo-svib,althoff-NiAl}
that there are large differences in the vibrational {\it entropies of
ordering\/} $S^{\rm vib}_{\rm ordered}-S^{\rm vib}_{\rm disord}$, which
should manifest themselves in shifts of the order-disorder transition
temperatures. There is another important class of thermodynamic
properties where the vibrational entropy may play a role, and which
has often been overlooked. Namely, it is the {\it entropy of
formation\/} with respect to the pure constituents, defined in analogy
with $\Delta H$ in Eq.~(\ref{eq:LDAenergies}):
\begin{eqnarray}
\nonumber
\Delta S_{\rm tot}^{\rm form} ({\rm A}_{1-x}{\rm B}_x, T) = 
S({\rm A}_{1-x}{\rm B}_x, T) \\
\label{eq:Sform}
- (1-x) S({\rm A}, T) - x S({\rm B}, T),
\end{eqnarray}
where $S({\rm A}, T)$ is the total entropy of the pure constituent 
$A$ at temperature $T$. It is often assumed that the
configurational entropy is the dominant contribution to 
$\Delta S_{\rm tot}^{\rm form}({\rm A}_{1-x}{\rm B}_x, T)$ 
because all other contributions
cancel out in Eq.~(\ref{eq:Sform}). The non-configurational entropy of
formation,
\begin{eqnarray}
\nonumber
\Delta S_{\rm non-conf}^{\rm form} ({\rm A}_{1-x}{\rm B}_x, T) = 
	\Delta S_{\rm tot}^{\rm form} ({\rm A}_{1-x}{\rm B}_x, T)\\
\label{eq:Snon-conf}
 - \Delta S_{\rm conf} ({\rm A}_{1-x}{\rm B}_x, T),
\end{eqnarray}
contributes to such important quantities as mutual
solubility limits and miscibility gap temperatures.

Noble metal alloys are excellent cases to test the values of $\Delta
S_{\rm non-conf}^{\rm form}$ since accurate experimental data on the
entropies of formation, $\Delta S_{\rm tot}^{\rm form}$, are
available, and the configurational entropy $\Delta S_{\rm conf}$ can
be calculated accurately using the thermodynamic integration technique 
described in Sec.~\ref{sec:simann}. Table~\ref{tab:Snon-conf} gives
the measured entropies of formation for disordered solid solutions
A$_{1-x}$B$_x$, $\Delta S_{\rm tot}^{\rm form}(x,T)$, the maximum
attainable configurational 
entropy $\Delta S_{\rm ideal}$, as well as the theoretically
calculated configurational entropy $\Delta S_{\rm conf}^{\rm calc}$,
and the derived value for the non-configurational entropy of formation, 
$\Delta S_{\rm non-conf}^{\rm form}$. It shows that the
size-mismatched noble metal systems have large amounts of $\Delta
S_{\rm non-conf}^{\rm form}$ in the disordered solid solution.
Since it is unlikely that these values of 
$\Delta S_{\rm non-conf}^{\rm form}$ are of electronic or magnetic
origin, we suggest that the excess entropy in the disordered solid
solutions of Ni-Au, Cu-Ag and Cu-Au is vibrational.
It is possible that the atomic relaxations lead to a softening of
lattice vibrations, although the physical mechanism of this softening
is unclear at present.

Sanchez {\it et al.\/}\cite{sanchez+moruzzi} in their study
of the Cu-Ag system noted that even a very crude model of the
vibrational entropy markedly improved the agreement with the
experimental solubility data. In the case of Ni-Au, which exhibits the
largest $\Delta S_{\rm non-conf}^{\rm form}$, it is possible to
reconcile the experimentally measured and theoretically calculated
miscibility gap 
temperatures only by taking into account the non-configurational entropy
of formation.\cite{cmw-niau}

The fact that Cu-Au also has a positive $\Delta S_{\rm non-conf}^{\rm
form}$ has little qualitative effect on the phase diagram since Cu and
Au are completely miscible from total energy and configurational
entropy considerations alone. Ag-Au is calculated to have a negative
$\Delta S_{\rm non-conf}^{\rm form}$, 
but its value is close to the experimental uncertainty in the
measurement of $\Delta S$.

\subsection{Bond lengths in random alloys}
\label{sec:bonds}

\begin{figure}
\epsfxsize=3.0in
\centerline{\epsffile{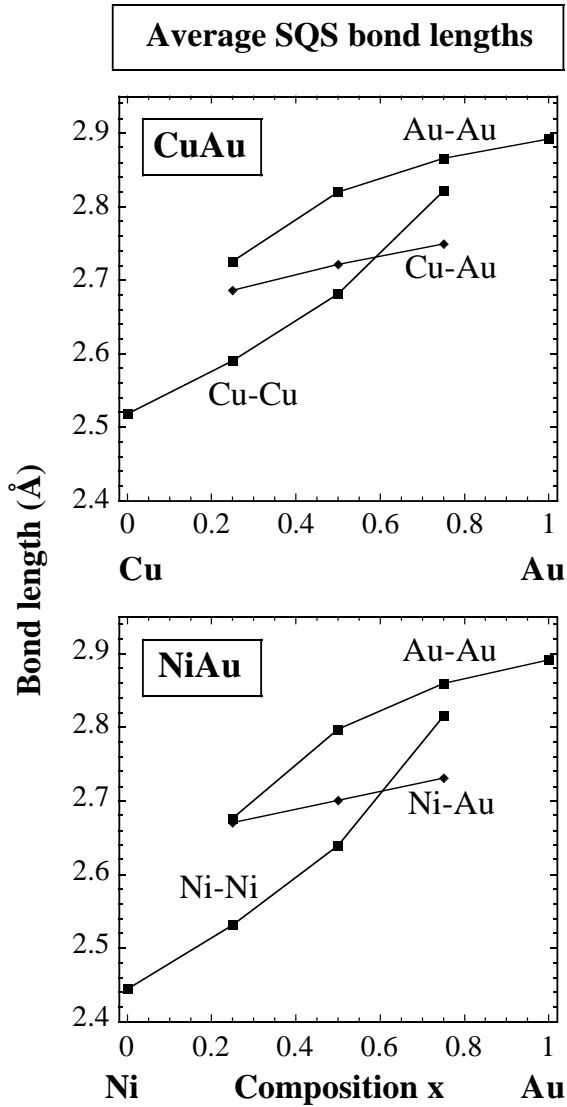}}
\vskip 5mm
\caption{SQS bondlengths for Cu-Au and Ni-Au.}
\label{fig:SQS_bonds}
\end{figure}

Since recent experimental measurements of the composition-dependence
of interatomic bond lengths in Cu-Au\cite{frenkel} and
Ni-Au\cite{exafs-niau} have found several unusual features, it is
interesting to address these trends from first-principles LDA
calculations. In the present work we model the atomic positions in the
random alloys using special
quasirandom structures\cite{SQS} (SQS). These periodic structures are
designed to reproduced the pair and multibody correlation functions
of the perfectly disordered configuration as closely as possible.
It has been shown\cite{SQS} that even small unit cell SQS's can
give rather accurate representation of the properties of random
alloys. We have performed LDA calculations for 8 atom/cell SQS's
at $x=\frac{1}{4}$ $(SQS14_a)$, $x=\frac{1}{2}$ $(SQS8_a, SQS8_b)$ 
and $x=\frac{3}{4}$ $(SQS14_b)$. The atomic positions and cell
coordinates have been fully relaxed to minimize the total energy.
The results for Cu-Au and Ni-Au interatomic bond lengths are shown in
Fig.~\ref{fig:SQS_bonds}. The main features are:

(i) In spite of the different phase diagram properties (Ni-Au phase
separates and Cu-Au orders at $T=0$~K), the calculated behavior of
bond lengths is very similar, which we attribute to the similar size
mismatch in both systems (12\% in Cu-Au and 15\% in Ni-Au).

(ii) Our calculations give three distinct bond lengths 
at all compositions, which is also observed
experimentally.\cite{exafs-niau,frenkel} Probably the most interesting
feature in Fig.~\ref{fig:SQS_bonds} is the crossing of
$R_{BB}(x)$ and $R_{AB}(x)$ curves at $x=\frac{3}{4}$ in both systems.
The measurements for Cu-Au\cite{frenkel} and
Ni-Au\cite{exafs-niau} indicate that this may indeed be correct,
since the deduced values around this composition are very close and
have large error bars. 

(iii) Another important feature, observed
experimentally and reproduced by our SQS results, is that $A-A$
bonds change much more as $x$ varies from 0 to 1 than 
$B-B$ bonds when $x$ varies from 1 to 0, suggesting that
the compressed bonds become increasingly stiff and the
expanded bonds weaken.
This behavior can be explained by the asymmetry in the
interatomic potential curves, which are rapidly hardening upon
compression and softening upon expansion. However,
our results for $R_{AA}$ at $x=\frac{3}{4}$ and 
$R_{BB}$ at $x=\frac{1}{4}$ are obtained from an average
of only 4 minority bonds in the $SQS14$ structures, and perhaps are 
not representative of a wider statistical sample.

(iv) It is interesting to note
that the predicted bond lengths between unlike atoms $R_{AB}$
do not follow the linear relation $R_{AB} = R_{AA} + x
(R_{BB}-R_{AA})$.

\section{Summary}
\label{sec:summary}

We have showed that accurate first-principles studies of alloys
with large size mismatches are now feasible using the mixed-space
cluster expansion method. This method has been applied to noble metal
alloys where vast amounts of experimental data and many
theoretical studies are available.

(i) The mixed-space cluster expansion has been generalized to include
the effects of nonlinear strain on the formation energies of
long-period superlattices. We find that the elastic energy, required to
lattice-match Cu and Ni to (100) surfaces of Au and Ag, is anomalously
low, leading to a very low constituent strain energy of (100)
superlattices. This effect is partly responsible for the stabilization
of new LDA ground states of Au-rich Cu-Au alloys.

(ii) In Au-rich Cu-Au, we predict new $T=0$~K ground states.
Our LDA results place $L1_2$ (CuAu$_3$), previously thought of as the
stable $T=0$ state of CuAu$_3$, higher in energy
than a family of superlattices along (100) direction. In particular,
MoPt$_2$-type CuAu$_2$ [Cu$_1$Au$_2$ superlattice along (100)] and a
complicated Cu$_1$Au$_4$Cu$_1$Au$_4$Cu$_1$Au$_2$Cu$_1$Au$_2$ (100)
superlattice are found to be the LDA ground states.

(iii) There are significant discrepancies (up to 50\%) between the
experimentally measured and calculated LDA mixing enthalpies
for Cu-Au alloys. This is surprising since the experimental mixing
enthalpies of Ni-Au and Ag-Au are reproduced very
well.\cite{zwlu-agau,cmw-niau}

(iv) The calculated order-disorder transition temperatures are in an
excellent agreement with experiment. For instance, 
$T_c^{\rm calc}(x=\frac{1}{4}) = 530$~K and
$T_c^{\rm calc}(x=\frac{1}{2}) = 660$~K, compared with
$T_c^{\rm expt}(x=\frac{1}{4}) = 663$~K and 
$T_c^{\rm expt}(x=\frac{1}{2}) = 683/658$~K.

(v) From the experimentally measured entropies of formation
$\Delta S^{\rm form}_{\rm tot}$ and the calculated configurational
entropies $\Delta S^{\rm calc}_{\rm conf}$, we obtain large
non-configurational (probably vibrational) entropies of formation in
the size-mismatched systems, $\Delta S^{\rm form}_{\rm non-conf}= 
\Delta S^{\rm form}_{\rm tot} - \Delta S^{\rm calc}_{\rm conf}$.
These entropies allow one to reconcile the experimental miscibility gap
temperature and formation enthalpies of Ni-Au with the theoretical LDA
values.\cite{cmw-niau}

(vi) Bond length distributions in Ni-Au and Cu-Au have been studied
via supercell calculations employing the special quasirandom
structure technique. The important qualitative features of recent
EXAFS measurements\cite{exafs-niau,frenkel} are correctly reproduced:
existence of distinct $A-A$, $B-B$ and $A-B$ bond lengths at all
compositions, possible crossing of $R_{AA}(x)$  and  $R_{AB}(x)$
around $x=\frac{3}{4}$ (where $x$ is the composition of the larger
constituent), softening of the shorter bond as $x \rightarrow 1$,
and deviations of the bond length $R_{AB}(x)$ between unlike atoms
from the linear Vergard's law.

\acknowledgements

This work has been supported by the Office of Energy Research,
Basic Energy Sciences, Materials Science Division, U.S. Department of
Energy, under contract DE-AC36-83CH10093.

\end{document}